\documentclass[11pt]{article}

\usepackage{color}
\usepackage{amsmath, amsfonts, amsthm, amssymb, graphicx}
\usepackage{xspace}
\usepackage{epsfig}
\usepackage{psfrag}
\usepackage{wrapfig}
\usepackage{hyperref}

\newcommand{\Complex}{\ensuremath{\mathbb{C}}\xspace}
\newcommand{\Real}{\ensuremath{\mathbb{R}}\xspace}

\newcommand{\Adst}{\ensuremath{\text{AdS}_3}\xspace}
\newcommand{\wAdst}{\ensuremath{\widehat{\text{AdS}}_3}\xspace}
\newcommand{\op}{\ensuremath{\mathcal{O}}\xspace}

\newcommand{\ket}[1]{\ensuremath{| #1 \rangle}\xspace}

\newcommand{\vev}[1]{\ensuremath{\langle #1 \rangle}\xspace}

\def\ie{{\it i.e.\ }}
\renewcommand{\Im}{\text{Im}}
 \renewcommand{\Re}{\text{Re}}

\let\a=\alpha \let\b=\beta   
    
\let\l=\lambda \let\m=\mu \let\n=\nu  \let\r=\rho

    \let\G=\Gamma

\newcommand{\be}{\begin{equation}}
\newcommand{\ee}{\end{equation}}
\def\ba{\begin{array}}
\def\ea{\end{array}}
\newcommand{\bea}{\begin{eqnarray}}
  \newcommand{\eea}{\end{eqnarray}}

\newcommand{\mx}{\mu}
\newcommand{\nx}{\nu}

\begin{document}

\begin{flushright}
ITF-2009-27
\end{flushright}
\begin{center}
\vskip 2cm
{\Large \bf Holography and wormholes in 2+1 dimensions\\}
\vskip 1cm
{\Large Kostas Skenderis${}^{a,b}$ and Balt C. van Rees${}^a$}\\
\vskip 0.7cm
{\it ${}^a$ Institute for Theoretical Physics,\\
P.O. Box 94485
1090 GL Amsterdam, The Netherlands}\\
\vskip 0.5cm
{\it ${}^b$ Korteweg-de Vries Institute for Mathematics\\
P.O. Box 94248
1090 GE Amsterdam, The Netherlands}  \\
{\tt K.Skenderis, B.C.vanRees@uva.nl}\\
\vskip 2cm

\end{center}

\begin{abstract}
\normalsize
We provide a holographic interpretation of a class of
three-dimensional wormhole spacetimes. These spacetimes have
multiple asymptotic regions which are separated from each other
by horizons. Each such region is isometric to the
BTZ black hole and there is non-trivial spacetime topology
hidden behind the horizons. We show that application of the
real-time gauge/gravity duality results in a complete
holographic description of these spacetimes with
the dual state capturing the non-trivial topology behind the horizons.
We also show that these spacetimes are in correspondence
with trivalent graphs and provide an explicit metric description
with all physical parameters appearing in the metric.
\end{abstract}

\newpage
\tableofcontents
\newpage
\section{Introduction and summary of results}

The gauge/gravity duality
\cite{Maldacena:1997re,Gubser:1998bc,Witten:1998qj} has significantly
enhanced our understanding of gravity and gauge theory.
This can be ascribed largely to a well developed dictionary
that translates results between string and gauge theory. Although the
entries in the dictionary are by now well understood for Euclidean
backgrounds, a \emph{real-time} dictionary along the lines of
\cite{Witten:1998qj} was developed only recently in \cite{us, us2}.
This real-time dictionary uses a construction that is
a holographic version of the closed time path method
of non-equilibrium QFT
\cite{Schwinger:1960qe,Bakshi:1962dv,Bakshi:1963bn,Keldysh:1964ud}
and results in a prescription that incorporates in the bulk
the information about the QFT initial and final
states via a Hartle-Hawking type construction
\cite{Hartle:1983ai,Maldacena:2001kr}. Thus this prescription
although originated from QFT considerations is also
in line with expectations from quantum gravity.

In this paper we apply the prescription of \cite{us,us2} to a class of
2+1-dimensional `wormhole' spacetimes that were found and studied in
\cite{Aminneborg:1997pz,Brill:1998pr}. Our main motivation is to
investigate global issues in gauge/gravity duality. Three
dimensional gravity is an ideal setup to study this problem
because of the absence of local degrees of freedom.
In the holographic context one finds that the general solution of the
bulk Einstein equations with a cosmological constant
in the
Fefferman-Graham gauge can be explicitly
obtained for general Dirichlet boundary
conditions specified by an arbitrary boundary metric
\cite{Skenderis:1999nb}.
In contrast to the higher
dimensional case, where in general the Fefferman-Graham expansion contains an
infinite number of terms, in three dimensions the series terminates
(see (\ref{eq:barefgmetric}))
and all coefficients can be expressed explicitly
in terms of the boundary metric and boundary stress energy
tensor. What is left to be done is to impose
regularity in the interior and this step requires global
analysis\footnote{Note also that
the Fefferman-Graham coordinates are
in general well-defined only in a neighborhood of the boundary
and they may not cover the entire spacetime.}.

\begin{wrapfigure}{R}{6cm}
\centering
\includegraphics[width=6cm,angle=270]{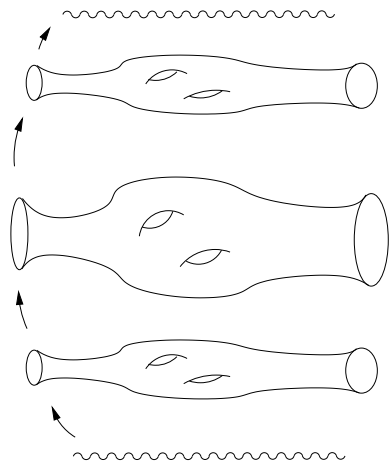}
\caption{\label{fig:sketch}A wormhole spacetime with two
outer regions corresponding to a Riemann surface of
genus 2 with 2 boundary components.}
\vskip 0.5 cm
\end{wrapfigure}

The wormholes are global solutions of 2+1 dimensional gravity with
a negative cosmological constant. They
can be thought of as generalized eternal BTZ black holes.
Whereas the spatial slices of an eternal BTZ black
hole have a cylindrical topology, in the wormholes the
spatial slices are general two-dimensional Riemann surfaces with boundary.
We sketch an example of a
wormhole in figure \ref{fig:sketch}.
These spacetimes have a number of different asymptotic regions,
which we will call outer regions in this paper, one for each boundary
component of the Riemann surface.
The outer regions are separated by horizons and there is a
non-trivial topology behind the horizons. The wormholes
are locally \Adst and should have a holographic
interpretation. Each outer region, however, is isometric to the
static BTZ black hole
and it would seem as though holographic data, which are obtained from the
behavior of the solution near the conformal boundary,
do not contain enough information to completely describe the wormhole
spacetime. This follows from a simple counting argument. The spacetimes
are uniquely determined given a Riemann surface of genus $g$ with $m$
boundaries. Such a Riemann surface is determined by $6 g - 6 + 3 m$ parameters.
Each of the outer regions however depends on only one parameter, the mass
of the BTZ black hole, so the holographic data from the $m$ outer regions
would seem to provide only $m$ parameters. We will shortly describe how
the real-time dictionary resolves this puzzle.

There are corresponding Euclidean solutions which have been discussed
in \cite{Krasnov:2000zq}. These spaces are \emph{handlebodies}, \ie
closed surfaces of genus $g$ filled in with hyperbolic three-space.
These are also generalizations of BTZ whose
Euclidean counterpart is a solid torus, \ie a handlebody of genus 1.
For these spacetimes a fairly straightforward application of the
Euclidean gauge/gravity prescription shows there is no
corresponding puzzle: the holographic
one-point function captures the non-trivial topology
and in particular does contain enough parameters to completely
describe these spaces. This indicates that it is
the real-time issues that are crucial in
understanding holography for the Lorentzian wormholes.

We will indeed find that once we properly apply the real-time
gauge/gravity prescription of \cite{us,us2} there is a direct and
unambiguous holographic interpretation of the entire Lorentzian
wormhole spacetimes.
The real-time prescription relies on gluing
to a given Lorentzian spacetime Euclidean spaces that
provide the initial and final states. A class of such Euclidean spaces
are the handlebodies described above, but we emphasize
that there are also other choices
one can make. Once the complete spacetime has been
specified (with the Euclidean parts representing initial/final states
included), the holographic one-point functions do carry enough information
about the spacetime and in particular the geometry behind
the horizons. This information is encoded
in the initial and final states.

The way this happens is instructive and reflects a number of subtle
points about the holographic dictionary. Recall that
because of the holographic conformal anomaly
\cite{Henningson:1998ey,Henningson:1998gx} the theory depends on the
specific boundary metric, not just its conformal
class. In particular, the expectation
value of the stress energy tensor changes anomalously under bulk
diffeomorphisms that induce a boundary Weyl transformation
\cite{deHaro:2000xn,Skenderis:2000in}.
Now as mentioned earlier, one can choose coordinates such that the
metric in any of the outer regions of the wormhole
is exactly that of the BTZ black hole. In these coordinates
the boundary metric is flat. According to the
prescription of \cite{us,us2}, however, the Lorentzian solution
should be matched in a smooth fashion to a corresponding Euclidean
solution. Euclidean solutions that satisfy all matching conditions
are provided by the handlebodies but these can never have a boundary metric
that is globally flat (because the Euler number of the
boundary Riemann surface is negative). One can arrange for
an everywhere smooth matching by performing a bulk diffeomorphism
on the Lorentzian side that induces an appropriate boundary
Weyl transformation such that the Lorentzian boundary metric now
matches with that of the handlebody.
This has the effect that the expectation value of the
stress energy tensor changes from its BTZ value to a new
value, which is
smooth as we cross from the Euclidean side to the Lorentzian side
(as it should be \cite{us2}). In other words, the initial
state via the matching conditions dictates a specific bulk
diffeomorphism on the outer regions of the Lorentzian
solution and as a result the holographic data extracted using
the solution in this coordinate
system encode the information hidden behind
the horizon.

Our results indicate that the dual state for a wormhole
with $n$ outer regions is an entangled state in a Hilbert
space that is the direct product of $n$ Hilbert spaces,
one for each component.
A reasonable guess for this state is that it is the state
obtained by the Euclidean path integral over the conformal
boundary of half of the Euclidean space glued at the
$t=0$ surface of the Lorentzian wormhole. This is a
Riemann surface with $n$ boundaries and in the case of the
handlebodies discussed above, it is precisely the Riemann
surface that serves as the $t=0$ slice of the wormhole.
If one traces out all components but one, then the reduced
description is given in terms of a mixed state in the remaining copy.

This paper is organized as follows. In the next section we
describe the wormhole spacetimes in detail and in section
\ref{sec:euclidean} and  we discuss the handlebodies. In sections
\ref{sec:holographic} and \ref{sec:Lorentzian} we discuss holography
for the handlebodies and the Lorentzian wormholes, respectively.
We emphasize that our analysis applies only to non-rotating wormholes.
The interesting possibility of extending the analysis to rotating wormholes
is discussed in section \ref{sec:remarks} along with several other
general remarks. We conclude in section
\ref{sec:conclusions} with an outlook.

In all of previous literature and in the main text of this paper
the wormhole spacetimes are described abstractly as quotients of a domain in
AdS$_3$. While this presents no loss of information, this description
is abstract and requires mastering prerequisite  mathematical
background in order to understand the properties of these
spacetimes. One should contrast this with the case of the
BTZ black hole \cite{Banados:1992wn,Banados:1992gq}
where one has an explicit metric containing the physical
parameters (the mass and angular momentum). The BTZ also has an
abstract representation as a quotient of a domain of AdS$_3$ but this
has not been used as much as the explicit metric description.
With the hope that a more concrete description of the
wormholes would make them more readily
accessible we derive in
appendix \ref{app:coordinatesystems} an explicit metric description where all parameters
that determine the spacetime appear in the metric and we summarize
this result here.

\begin{figure}
\centering
\psfrag{M1}{$M_1$}
\psfrag{M2}{$M_3$}
\psfrag{M3}{$M_4$}
\psfrag{M4}{$M_5$}
\psfrag{M5}{$M_6$}
\psfrag{M6}{$M_7$}
\psfrag{M7}{$M_2$}
\psfrag{x2}{$\chi_3$}
\psfrag{x3}{$\chi_4$}
\psfrag{x4}{$\chi_5$}
\psfrag{x5}{$\chi_6$}
\psfrag{x6}{$\chi_7$}
\includegraphics[width=10cm]{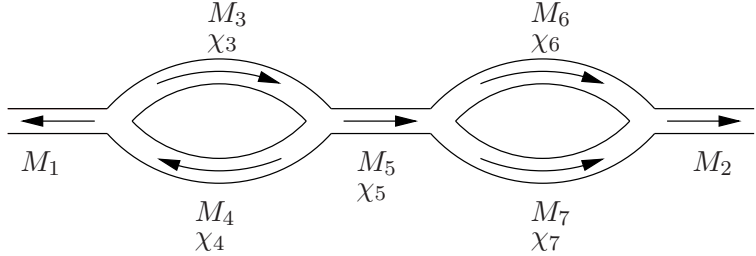}
\caption{\label{fig:graph}A fatgraph representing the
wormhole spacetime sketched in figure \ref{fig:sketch}.}
\end{figure}

All information about the wormhole can be summarized in an oriented
trivalent fatgraph, like the one in figure \ref{fig:graph}.
For a wormhole that is based on a Riemann surface of
genus $g$ with $m$ boundaries, this graph should have $m$
outer edges (ends) and
$(3 g - 3 + m)$ inner edges. With every outer edge
we associate one parameter $M_k$
and with every inner edge
two parameters $M_i, \chi_i$,
where $k{=}1,\dots,m$ and $i{=}m+1, \ldots, 3g{-}3{+}2m$.
This yields a total of $(6 g -6 + 3 m)$  parameters, which is indeed the
correct number of moduli for a Riemann surface of genus $g$ with
$m$ boundaries\footnote{As we review in the appendix,
the parameters $\{M_I, \chi_i\} \ (I=k,i)$ are directly related to the
Fenchel-Nielsen coordinates of the moduli space of the Riemann surface.}.
We now associate a coordinate chart for every edge
of the fatgraph and every such chart has a canonical metric on it.
The precise definition of the coordinate charts as well as the
meaning of the orientation is given in the appendix.
To complete the description we need to
specify the transition functions in the overlap regions and
these are also given in the appendix.

Thus, the spacetime is described by the graph and two different metrics, one
for the outer charts and one for the inner charts. The metric in the $k$th outer chart takes the form:
\be
\label{out}
ds^2_k = \frac{\rho^2 + M_k}{\cosh^2 (\sqrt{M_k} \tilde \tau)}(-d\tilde \tau^2 + d\varphi^2) + \frac{d\rho^2}{\rho^2 + M_k}\,.
\ee
The corresponding $(\tilde \tau, \rho,\varphi)$ coordinate system has coordinate ranges,
\be
\tilde \tau \in \Real\,,\qquad \qquad \varphi \sim \varphi + 2\pi\,,\qquad \qquad \frac{\cosh(\sqrt{M_k} \tilde \tau)\r}{\sqrt{\r^2 + M_k}} > - \frac{\b^2}{1 + \b^2}\,,
\ee
where $\b$ is defined in the appendix. These coordinates extend beyond the future and past horizons, which lie at
\be
\r = \sqrt{M_k} |\sinh(\sqrt{M_k} \tilde \tau)|\,.
\ee
If we restrict ourselves to the region outside of the horizons we may also put the metric in the static BTZ form,
\be \label{outbtz}
ds_k^2 = - (r^2 - M_k) dt^2 + \frac{dr^2}{r^2 - M_k} + r^2 d\phi^2\,,
\ee
with coordinate ranges, $r > M_k$, $t \in \Real$ and $\phi \sim \phi + 2 \pi$. In these metrics $M_k$ is the parameter of the corresponding outer edge.
The metric in the $i$th inner chart is given by
\be
\begin{split} \label{inn}
ds_i^2 &= \frac{1}{\cosh^2(t)}
\Big( -dt^2 + \frac{\m_i^2 dr^2}{(\m_i r + \n_i)^2 + \cos^2(\chi_i)}
+ M_i \big(1 + (\m_i r + \n_i)^2\big) d \psi^2 \\
&\qquad \qquad
- \frac{2 \mu_i \sqrt M_i \sin(\chi_i)}{\sqrt{(\m_i r + \n_i)^2
+ \cos^2 (\chi_i)}} \,
d\psi d r \Big)\,,
\end{split}
\ee
with coordinate ranges, $ r \in [-1,1],\ \tau \in \Real$ and
$\psi \sim \psi + 2\pi$.
This is a time-dependent metric of constant negative curvature
which (as far as we know) has not
appeared before in the literature.
The parameters $M_i$ and $\chi_i$ are the parameters associated with the
$i$th inner edge. The parameters $\m_i, \n_i$ on the other hand
are functions of the $M$ parameters, see the discussion in
section \ref{sec:parameters}.
Note that both metrics (\ref{out}) and (\ref{inn}) have
a $U(1)$ isometry, the transition functions however do not respect this
symmetry and the entire spacetimes is not $U(1)$ symmetric.

\section{Lorentzian wormholes}
\label{sec:lorentzian}
In this section we describe the Lorentzian wormholes. We show how they can be obtained as quotients of a part of \Adst and discuss their physical properties. The material in this section summarizes discussions in  \cite{Aminneborg:1997pz,Brill:1998pr,Barbot:2005fv,Barbot:2005qk,mythesis}. We will occasionally use results from Teichm\"uller theory; more information on this topic can be found in \cite{imayoshitaniguchi,lehto,nag,mythesis}.

\subsection{Wormholes as quotient spacetimes}
The wormholes are obtained as follows. One starts with a Riemann surface $S$ which is a quotient of the upper half plane $H$ with respect to some discrete subgroup $\Gamma$ of $SL(2,\Real)$. The upper half plane is then embedded into \Adst and the action of $\Gamma$ is extended to \Adst entirely. After removing certain regions in \Adst that would lead to pathologies, one may take the quotient of the remainder with respect to $\Gamma$, which will give us the wormhole spacetime we are after. The topology of such a spacetime is $S \times \Real$, with $S$ the Riemann surface we started with and $\Real$ the time direction. The aim of this subsection is to discuss this procedure in more detail.

\subsubsection{Riemann surfaces}

Consider a \emph{Riemann} surface $S$ with $m > 0$ circular boundaries
but no punctures\footnote{Recall that a Riemann surface is a topological
two-dimensional surface equipped with a complex structure. One can distinguish
between punctures and circular boundaries precisely because of the
complex structure.}. As follows from the uniformization theorem, such a
Riemann surface can be described as a quotient of the upper half plane
$H$ by some discrete subgroup $\Gamma$ of $PSL(2,\Real)$:

\be
S = H/\Gamma\,,
\ee
where the action of
\be
\begin{pmatrix} a & b \\ c & d \end{pmatrix} \in PSL(2,\Real) \equiv SL(2,\Real)/\{\pm \mathbf 1\}
\ee
on $H$ is given by
\begin{equation}
\label{eq:mobius}
z \mapsto \frac{az + b}{cz +d}\,.
\end{equation}
Since these transformations act as isometries for the standard negatively curved metric on $H$,
\be
\label{eq:metriconh}
ds^2 = \frac{dz d\bar z}{\text{Im}(z)^2}\,,
\ee
this metric descends to a metric on $S$. Up to a constant rescaling, this is the unique hermitian metric of constant negative curvature on $S$ (given the complex structure of $S$) and $\Gamma$ is unique up to conjugation. We shall require absence of conical singularities on $S$, which means that the nontrivial elements of $\Gamma$ cannot have fixed points in $H$. A simple analysis of the fixed points of \eqref{eq:mobius} tells us that we should require that for all elements $\gamma \in \Gamma$ we have
\be
|a+d| \geq 2\,.
\ee
Furthermore, absence of any punctures on $S$ translates into $|a+d| > 2$ for all nontrivial $\gamma$. We then say that $\Gamma$ consists of only \emph{hyperbolic} elements (and the identity), and we call it a \emph{Fuchsian group of the second kind}.

\begin{figure}
\centering
\includegraphics[width=9cm]{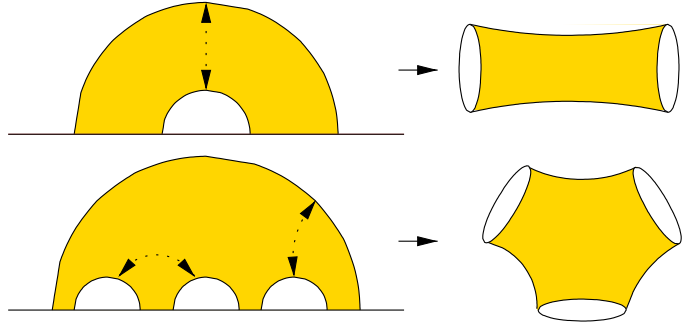}
\caption{\label{fig:fundamentaldomain}On the left we sketched two fundamental domains in $H$. The boundaries are pairwise glued together as indicated by the arrows. After the gluing we find the Riemann surfaces shown on the right.}
\end{figure}

A particularly convenient way to visualize $S$ as a quotient of $H$ is to define a \emph{fundamental domain} in $H$, basically a domain in $H$ whose boundary in $H$ consists of various segments that are pairwise identified by generators of $\Gamma$. For convenience we may take these segments to be geodesic segments, which are circular arcs in $H$. Two examples of a fundamental domain are sketched in figure \ref{fig:fundamentaldomain}.

From the theory of Fuchsian groups we obtain that the fixed points of such a group $\Gamma$ form a nowhere dense subset of the conformal boundary $\partial H$ of $H$, which is the real line plus a point a infinity. We will call this set the \emph{limit set} and denote it as $\Lambda(\Gamma)$. Notice that $\Lambda(\Gamma)$ is invariant under the action of $\Gamma$.

\subsubsection{\Adst}
To find the wormhole spacetime associated to $S$, we first fix some coordinate systems and conventions for \Adst. We define \Adst
as the surface
\be
- U^2 - V^2 + X^2 + Y^2 = -1\, ,
\ee
in $\Real^{2,2}$, where the metric has the form
\be
ds^2 = -dU^2 - dV^2 + dX^2 + dY^2\,,
\ee
and we have set the AdS radius $\ell^2 = 1$.
By combining $(U,V,X,Y)$ into a matrix,
\be
\begin{pmatrix}
V + X &  Y + U\\
Y - U & V - X
\end{pmatrix}\,,
\ee
we may identify the hyperboloid with the space of real unit determinant matrices, \ie the group $SL(2,\Real)$. The connected component of the identity of the isometry group of \Adst,
\be
\text{Isom}_0(\Adst) = (SL(2,\Real) \times SL(2,\Real))/{\mathbb Z_2}\,,
\ee
acts by left and right multiplication: if $(\gamma_1,\gamma_2)\in SL(2,\Real)\times SL(2,\Real)$ then their action on \Adst is defined by
\be
\label{eq:adstaction}
\begin{pmatrix}
V + X &  Y + U\\
Y - U & V - X
\end{pmatrix}
\mapsto
\gamma_1
\begin{pmatrix}
V + X &  Y + U\\
Y - U & V - X
\end{pmatrix}
\gamma_2^T\,.
\ee
Taking the transpose of $\gamma_2$ is a convention which will turn out to be convenient below.

We may describe a patch in the hyperboloid with Poincar\'e coordinates $(t,x,y)$ defined by
\be
\label{eq:poincarecoords}
t = \frac{U}{V-X}\,, \qquad x = \frac{Y}{V-X}\,, \qquad y = \frac{1}{V-X}\,.
\ee
In these coordinates, the metric takes the form
\be
ds^2 = \frac{-dt^2 + dx^2 + dy^2}{y^2}\,.
\ee
Although the Poincar\'e coordinate system may not cover the entire region of interest, the coordinate horizon at $y \to \infty$ will not be important in what follows.

\subsubsection{Constructing wormholes}
We can now construct a three-dimensional wormhole spacetime from the Riemann surface $S = H/\Gamma$. We begin by extending the action of the isometries of $H$ to isometries on \Adst via the homomorphism:
\be
PSL(2,\Real) \hookrightarrow (SL(2,\Real) \times SL(2,\Real))/{\mathbb Z_2}\,,
\ee
which is given explicitly by $PSL(2,\Real) \ni \gamma \mapsto (\gamma,\gamma) \in SL(2,\Real) \times SL(2,\Real)$. One may check that elements of the form $(\gamma,\gamma)$ leave the slice $U = 0$ invariant when they act on \Adst according to \eqref{eq:adstaction}. Furthermore, their action on the slice $U=0$ is exactly of the form \eqref{eq:mobius} when we define $z = x + iy$ with $(x,y)$ the Poincar\'e coordinates on this slice.

The image of $\Gamma$ under this homomorphism is a discrete subgroup of Isom$_0(\Adst)$ which is isomorphic to $\Gamma$ and which we denote as $\hat \Gamma$. One may now try to take a quotient like $\Adst/ \hat \Gamma$, which clearly contains $S = H/\Gamma$ as the slice given by $U=0$. However, away from the slice $U=0$ this quotient turns out to have closed null or timelike curves. To get a spacetime free of pathologies we proceed as follows.

The embedding of $H$ in AdS$_3$ as the slice $U=0$ can be directly extended to an embedding of $\partial H$ in the conformal boundary of AdS$_3$. This extension maps the limit set $\Lambda(\Gamma)$ to a subset of the conformal boundary of AdS$_3$, which we denote as $\Lambda(\hat \Gamma)$. We then pass to the universal covering space of the hyperboloid and \emph{remove} from it all points with a timelike or lightlike separation to $\Lambda(\hat \Gamma)$ (after a standard conformal rescaling of the metric that brings the radial boundary to finite distance). Informally speaking, we are removing the filled forward and backward semi-lightcones emanating from every point in $\Lambda(\hat \Gamma)$. We call the remainder \wAdst which notably includes the original slice $U=0$ entirely. The elements of $\hat \Gamma$ leave $\Lambda(\hat \Gamma)$ invariant and, being isometries, they map lightcones to lightcones so they also leave $\wAdst$ invariant. Furthermore, the quotient
\be
M = \wAdst/\hat \Gamma
\ee
is a spacetime that is free of closed timelike curves and conical singularities \cite{Barbot:2005fv,Barbot:2005qk} and contains $S = H/\Gamma$ as a hypersurface. These spacetimes are what we call the $2+1$-dimensional wormholes.

\subsection{Physical properties}

We briefly discuss some physical properties of the wormholes. First of
all, they are of course locally \Adst but, as was mentioned above,
their global topology is of the form $S \times \Real$ with $S$ a
surface with $m > 0$ circular boundaries and $\Real$ representing
time. We sketched an example in figure \ref{fig:sketch}, where $S$ has
genus 2 and has 2 boundary components. The wormholes can have an
arbitrary number $m > 0$ cylindrical boundaries, and $S$ can have
arbitrary genus $g \geq 0$. There are two special cases: when $m=2$
and $g=0$ we obtain the eternal static BTZ black hole and the case
$m=1$, $g=0$ is just AdS.

Except for the eternal BTZ black hole described already in
\cite{Banados:1992gq}, none of the wormholes have globally defined
Killing vector fields since no such isometry of \Adst
commutes with all the elements in $\Gamma$. On the other hand, all
wormholes admit a discrete $\mathbb Z_2$ isometry, which acts as time
reflection $U \leftrightarrow -U$ and therefore leaves the $U=0$ slice
invariant. The wormholes are not geodesically complete and begin with
and end on locally Milne-type singularities. Furthermore, these
singularities have associated black and white hole horizons (not drawn
in figure \ref{fig:sketch}).

Perhaps surprisingly, the $m$ segments of the spacetime between the
horizons and the conformal boundaries are \emph{exactly} the same as
for the BTZ black hole \cite{Aminneborg:1997pz}.
More precisely, we find that these segments
can be covered by a $(t,r,\phi)$ coordinate system with the coordinate
ranges $r > M$, $t \in \Real$ and $\phi \sim \phi + 2 \pi$, in which
the metric is of the form
\be
\label{eq:btzmetric}
ds^2 = - (r^2 - M) dt^2 + \frac{dr^2}{r^2 - M} + r^2 d\phi^2\,.
\ee
The mass $M$ can be different for the $m$ different boundaries, but it
should always be strictly positive so we do not `pinch off' the rest
of the wormhole. We will call these $m$ segments the \emph{outer
  regions} of the wormhole, and what remains when we excise these
segments we call the \emph{inner region}.
Notice that what we call the outer region is precisely the domain of
outer communication \cite{Aminneborg:1997pz}. What was called the
`exterior region' in \cite{Aminneborg:1997pz} is obtained by keeping
only the region outside of the future horizon, but we will never
consider this region here. The fact that the nontrivial topology is hidden
behind the horizons is in agreement with the general discussion of
\cite{Galloway:1999bp}.

Depending on the genus of
$S$, the geometry in the inner region is specified by a discrete
number of parameters, namely the moduli of $S$. One may for example
think of these parameters as the elements $(a_i,b_i,c_i,d_i)$ of a set
$\{\gamma_i\}$ of generators of $\Gamma$. It will be important for
what follows to notice that these parameters do not show up in the
metric on the outer regions if we put the metric in the form
\eqref{eq:btzmetric}. On the other hand, in appendix
\ref{app:coordinatesystems} we present a set of different coordinate
systems that can be used to describe the wormholes as well. In these
coordinate systems the coordinate ranges are natural and
the metric features several parameters that are
geometric (rather than abstract matrix elements). For example, some of
the parameters are directly related to the lengths of certain cycles
on the surface. As we explain in more detail in the appendix, the
combination of all parameters from the different charts that make up
the surface can be used to completely describe the
spacetime.\footnote{The parameters $(a_i,b_i,c_i,d_i)$ are similar to
  the \emph{Fricke} coordinates on the moduli space of $S$, whereas
  the metric we find in appendix \ref{app:coordinatesystems} features
  parameters that are similar to \emph{Fenchel-Nielsen} coordinates on
  the moduli space of $S$. These coordinate systems on the moduli or
  rather Teichm\"uller space of $S$ are described in more detail in
  for example \cite{imayoshitaniguchi}.}

It is straightforward to embed the wormholes into string theory, since the wormholes are locally just \Adst. For example, a wormhole times $S^3 \times T^4$ with a constant dilaton and three-form flux is an asymptotically locally \Adst solution of type IIB supergravity. However, these solutions are not supersymmetric.

\section{Euclidean wormholes}
\label{sec:euclidean}

In this section we describe `Euclidean wormholes'. These Euclidean spaces
are \emph{handlebodies} and one may think of them as closed Riemann
surfaces filled in with three-dimensional hyperbolic space. They are
a natural generalization of  the Euclidean BTZ black hole, which is
a solid torus \cite{Carlip:1994gc}. These spaces were considered first in a
holographic context in \cite{Krasnov:2000zq}, where it was argued that
they are natural Euclidean analogues of the Lorentzian wormholes,
even though they are not obtained by analytic continuation of
a globally defined time coordinate. We will see later that they are
indeed suitable Euclidean counterparts of the Lorentzian wormholes,
in the sense of
the real-time gauge/gravity prescription of \cite{us,us2},
but we will also show that they are not the only possible
Euclidean counterparts.

\subsection{Construction}
We will again describe the handlebodies via a quotient construction.
Recall that  Euclidean (unit radius)  \Adst, denoted by $H^3$,
is defined as the hyperboloid
\be
U^2 - V^2 + X^2 + Y^2  = -1\,,
\ee
with $V > 0$ in $\Real^{1,3}$ with the metric:
\be
ds^2 = dU^2 - dV^2 + dX^2 + dY^2\,.
\ee
We may again combine $(U,V,X,Y)$ into a matrix:
\be
\begin{pmatrix}
V + X &  Y + iU\\
Y - iU & V - X
\end{pmatrix}\,
\ee
which maps $H^3$ into the space of hermitian unit determinant matrices. An element $\gamma$ of the connected component of the identity of the isometry group of $H^3$,
\be
\text{Isom}_0(H^3) = PSL(2,\Complex)\,,
\ee
acts on $H^3$ as
\be
\label{eq:eadsaction}
\begin{pmatrix}
V + X &  Y + i U\\
Y - i U & V - X
\end{pmatrix}
\mapsto
\gamma
\begin{pmatrix}
V + X &  Y + i U\\
Y - i U & V - X
\end{pmatrix}
\gamma^\dagger\,.
\ee
Notice that $PSL(2,\Complex)$ maps the upper hyperboloid to itself.

We may again define Poincar\'e coordinates $(\tau,x,y)$ via
\be
\label{eq:euclpoinc}
\tau = \frac{U}{V-X}\,, \qquad x = \frac{Y}{V-X}\,, \qquad y = \frac{1}{V-X}\,.
\ee
In these coordinates, the metric takes the form
\be \label{eq:euclPoicmet}
ds^2 = \frac{d\tau^2  + dx^2 + dy^2}{y^2}\,.
\ee
This time there are no coordinate singularities and this metric covers all of $H^3$.

To find the Euclidean analogue of the wormholes, we again start with the Riemann surface $S= H/\Gamma$. The action of $\Gamma$ on $H$ can again be extended to an action on $H^3$ entirely, this time via the trivial homomorphism
\be
PSL(2,\Real) \hookrightarrow PSL(2,\Complex)\,,
\ee
(\ie any element of $PSL(2,\Real)$ is also an element of $PSL(2,\Complex)$).
One may again check that real elements in $PSL(2,\Complex)$ leave the slice $U=0$ invariant when they act on $H^3$ according to \eqref{eq:eadsaction}. Furthermore, their action on the slice $U=0$ is again of the form \eqref{eq:mobius} if we define $z = x + iy$ with $(x,y)$ the Poincar\'e coordinates on this slice.

After using this homomorphism to map $\Gamma$ to $\hat \Gamma$ in Isom$_0(H^3)$, we can define the quotient
\be
M_e = H^3/\hat \Gamma\,,
\ee
which now never leads to pathologies; $M_e$ is a smooth and geodesically complete manifold. This quotient again contains $S = H/\Gamma$ as the $U=0$ slice, and $M_e$ also admits a $\mathbb Z_2$ isometry that leaves this surface invariant.

Let us now show why we call $M_e$ a handlebody. We can extend the action of $\hat \Gamma$ to the conformal boundary of $H^3$ which is an $S^2$. Consider an element $\gamma$ of $\hat \Gamma$, \ie a real element of $PSL(2,\Complex)$, acting as \eqref{eq:mobius} on the $U = 0$ slice. Its extension to $H^3$ entirely is found most easily by noticing that, according to \eqref{eq:eadsaction}, real elements of Isom$_0(H^3)$ leave slices of constant $U = \tau/y$ invariant and act on these slices exactly as on the slice $U=0$. In the limit where $y \to 0$, we recover the action of $\gamma$ on the conformal boundary, which is just the same as on the slice $U=0$\,,
\be
\label{eq:actions2}
\gamma: w \mapsto \frac{aw + b}{cw + d}\,,
\ee
but this time with $w = x + i\tau$.

\begin{figure}
\centering
\psfrag{t}{$\tau$}
\psfrag{x}{$x$}
\includegraphics[width=13cm]{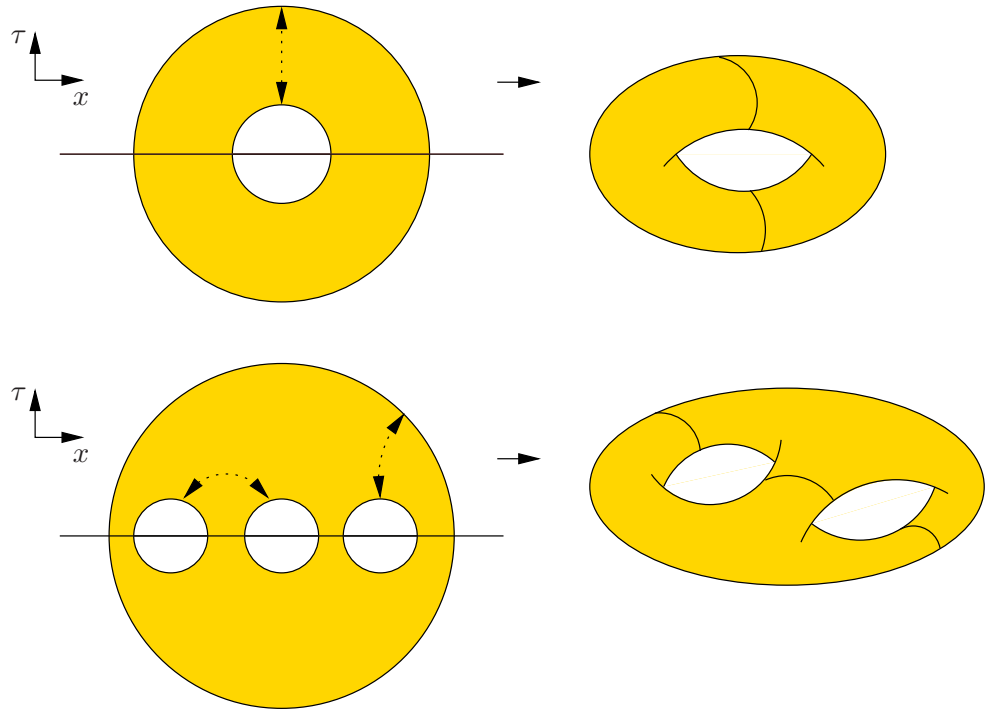}
\caption{\label{fig:schottkyfundamental}The Schottky double of the Riemann surfaces of figure \ref{fig:fundamentaldomain} is constructed by gluing two copies of the fundamental domain to each other and identifying the boundaries. The line $\tau = 0$ is invariant and the Schottky double surface is symmetric under reflection in this line. The limit set $\Lambda(\hat \Gamma)$ is a subset of the line $\tau = 0$ but is not shown here. It has to be removed from the $(\tau,x)$ plane before taking a quotient.}
\end{figure}

From \eqref{eq:actions2} we find that the great circle $\tau = 0$ is invariant because $a,b,c,d$ in \eqref{eq:actions2} are all real. Just as in the Lorentzian case, this circle contains the limit set $\Lambda(\hat \Gamma)$. After removing the limit set, the quotient of the remainder $S^2\backslash \Lambda(\hat \Gamma)$ with respect to $\hat \Gamma$ is a smooth manifold. As can be seen from figure \ref{fig:schottkyfundamental}, it consists of two copies of $S$, one from the upper and one from the lower half plane, glued together along their $m$ boundaries. This surface is called the \emph{Schottky double} $S_d$ of $S$. If $S$ has genus $g$ and $m$ holes, then $S_d$ has genus $2g + m -1$ and no holes. Since $S_d$ is just the conformal boundary of $M_e$, we may think of $M_e$ as a filled $S_d$. This shows that $M_e$ is indeed a handlebody.

\begin{figure}
\centering
\psfrag{t}{$\tau$}
\psfrag{x}{$x$}
\psfrag{y}{$y$}
\includegraphics[width=10cm]{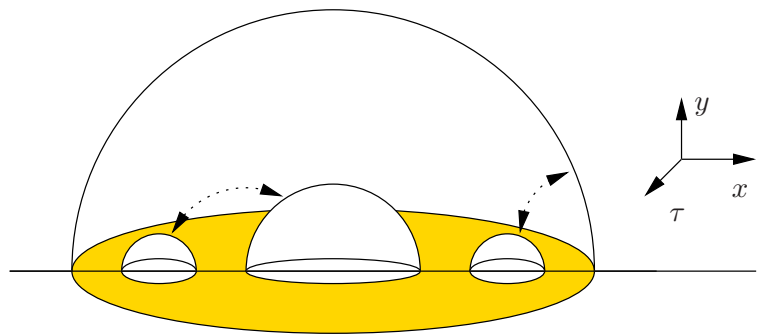}
\caption{\label{fig:handlebodyfundamental}The extension of the fundamental domain for $\hat \Gamma$ from the $S^2$ to $H^3$ is bounded by a set of hemispheres that should be pairwise identified. We recover $S$ as the surface given by $\tau = 0$.}
\end{figure}

A fundamental domain for $M_e$ in $H^3$ is sketched in figure \ref{fig:handlebodyfundamental} and can be found by extending the circles on the boundary $S^2$ to hemispheres in $H^3$. The fundamental domain for the original surface $S$ is then embedded in this three-dimensional fundamental domain as the surface given by $\tau = 0$.

\section{Holographic interpretation of Euclidean wormholes}
\label{sec:holographic}

We discuss in this section the holographic interpretation
of the Euclidean wormholes. Our discussion, which builds
on \cite{Krasnov:2000zq, Krasnov:2001cu, Krasnov:2003ye,mythesis},
is a fairly straightforward application of Euclidean holography.
In the next section we will turn to Lorentzian wormholes,
where things are more subtle.

Recall that the boundary $S_d$ of the handlebody is a closed
Riemann surface with $g > 1$ and therefore naturally has a metric of
constant negative curvature. Below, following
\cite{Krasnov:2001cu,mythesis}, we holographically compute the
one-point function of the stress energy tensor for this background
metric.

\begin{figure}
\centering
\psfrag{Gh}{$\hat \Gamma$}
\psfrag{Gd}{$\Gamma_d$}
\psfrag{J}{$J$}
\psfrag{Sd}{$S_d$}
\includegraphics[width=10cm]{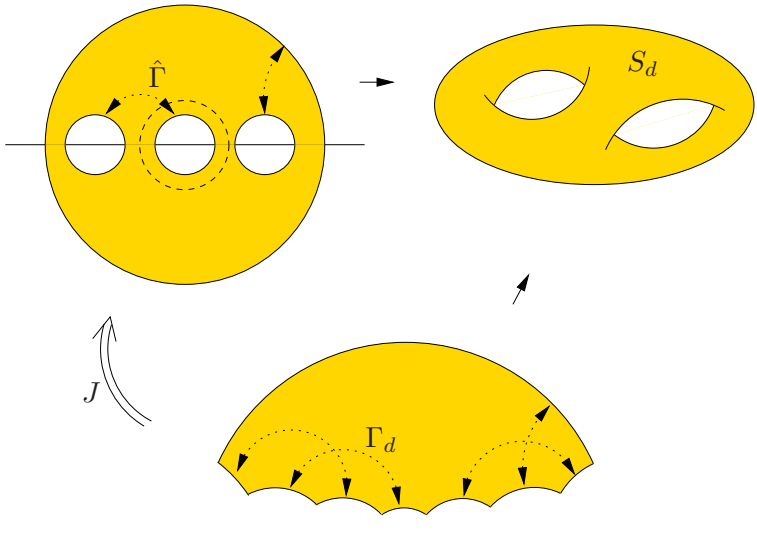}
\caption{\label{fig:J}The Riemann surface $S_d$ was originally obtained as $(S^2 \backslash \Lambda(\hat \Gamma))/\hat \Gamma$. However, like any closed Riemann surface with $g > 1$ it can also be described as $H/\Gamma_d$ for some $\Gamma_d$ for which we have drawn a fundamental domain in the bottom figure. $J$ is a locally biholomorphic map interpolating between the two descriptions. The dashed circle is a homotopically nontrivial closed curve on $S_d$ that can be contracted in the bulk.}
\end{figure}

The negative curvature metric on $S_d$ is obtained by describing $S_d$
as a quotient of $H$, that is $S_d = H / \Gamma_d$. Above we described
$S_d$ as a quotient of the conformal boundary $S^2$ of $H^3$, with the
limit set $\Lambda(\hat \Gamma)$ removed, with respect to the group
$\hat \Gamma$, that is $S_d = (S^2 \backslash \Lambda(\hat
\Gamma))/\hat \Gamma$. As sketched in figure \ref{fig:J}, this is just
a different description of the same Riemann surface. Therefore, there
should be a locally biholomorphic map $J : H \to S^2$ between the two
descriptions. Such a map should be compatible with the actions of
$\Gamma_d$ and $\Gamma$, in the
sense that for every $\gamma_d \in \Gamma_d$ there should exist a $\gamma
\in \hat \Gamma$ such that $J \circ \gamma_d = \gamma \circ J$.
Now consider the case where
$\gamma$ is trivial for a nontrivial $\gamma_d$. Since $\gamma_d$
corresponds to a nontrivial one-cycle on $S_d$, the image of this
one-cycle under $J$ must be a nontrivial closed curve on $S^2
\backslash \Lambda(\hat \Gamma)$. The only way to do this is to let
this curve encircle a nonempty subset of $\Lambda(\hat \Gamma)$ on the
$S^2$, but such a one-cycle \emph{is} contractible in the bulk
manifold. For example, the dashed circle drawn within the fundamental
domain of figure \ref{fig:J} can be continuously
shrunk to a point by moving it inside the bulk, as can be seen from
figure \ref{fig:handlebodyfundamental}. Therefore, precisely those
$\gamma_d$ for which $J \circ \gamma_d = J$ correspond to contractible
cycles in the bulk. The map $J$ thus determines the filling of $S_d$:
different maps $J$ (up to composition with an element of
$PSL(2,\Real)$ or $PSL(2,\Complex)$) precisely correspond to the
different fillings of $S_d$.
It therefore suffices to know $J$ in order to
know which cycles of $S_d$ are filled to give a handlebody and
therefore to determine the Euclidean bulk geometry.

Since $J$ is by construction locally biholomorphic, we can use the
locally defined $J^{-1}$ to pull back the metric \eqref{eq:metriconh}
from $H$ to $S^2 \backslash \Lambda(\Gamma)$. In Poincar\'e
coordinates for $H^3$ defined in \eqref{eq:euclpoinc}, the induced
metric on the boundary $S^2$ was the flat metric $ds_{(0)}^2 = dw
d\bar w$ with $w = x + i\tau$. On the other hand, when we pull back
the metric from $H$ using $J^{-1}$, we find a metric on this $S^2$
which is of the form:
\be
\label{eq:metricons2}
ds_{(0)}^2 = \Big|\frac{dJ^{-1}}{dw}\Big|^2 \frac{dw d\bar w}{\text{Im}(J^{-1}(w))^2} \equiv e^{2\sigma}dw d\bar w\,.
\ee
This metric is just a Weyl rescaling of the original metric:
\be
\label{eq:weylmetric}
dw d\bar w \mapsto e^{2\sigma}dw d\bar w\,,
\ee
where we note that $\sigma$ becomes singular whenever $\Im(J^{-1}(w))$ vanishes, which is precisely at the fixed point set $\Lambda(\hat \Gamma)$ on $S^2$.

We may now investigate what happens to the one-point function of the
stress energy tensor. Recall that in three dimensions the
metric near the conformal boundary can always be put in
the Fefferman-Graham form \cite{Skenderis:1999nb},
\be
\label{eq:barefgmetric}
ds^2  = \frac{d\rho^2}{\rho^2}+ \frac{1}{\rho^2}(g_{(0)ij} + \rho^2 g_{(2)ij} + \rho^4 g_{(4)ij}) dx^i dx^j, \qquad g_{(4)ij} = \frac{1}{4} (g_{(2)} g_{(0)}^{-1} g_{(2)})_{ij}\,,
\ee
and  the one-point function of the stress energy tensor
in the dual state is given by \cite{deHaro:2000xn}
\be
\label{eq:vevTgeneral}
\vev{T_{ij}} = 2 g_{(2)ij} + R_{(0)}g_{(0)ij}\,,
\ee
with $R_{(0)}$ the scalar curvature of $g_{(0)ij}$ and we set $16 \pi G_N = 1$.

In the case at hand, starting with the bulk metric
\eqref{eq:euclPoicmet}, we find that $ds^2_{(0)} = dw d\bar w$ and
$\vev{T_{ij}}_{g_{(0)}} = 0$. A bulk diffeomorphism that induces the
Weyl rescaling in \eqref{eq:weylmetric} has the effect of transforming the
$g_{(2)}$ such that \cite{Skenderis:2000in}
\be
\vev{T_{ww}}_{e^{2\sigma} g} = \vev{T_{ww}}_{g} + 2 \partial_w^2 \sigma - 2 (\partial_w\sigma)^2\,,
\ee
in agreement with CFT expectations. Since in our case
\be
\sigma = \frac{1}{2} \ln ( \partial_w J^{-1} ) + \frac{1}{2} \ln(\overline{\partial_w J^{-1}})  - \ln \Big(\frac{1}{2i}(J^{-1} - \bar J^{-1})\Big)\,,
\ee
we obtain directly that
\be
\label{eq:Tons2}
\vev{T_{ww}}_{e^{2\sigma}g} = \frac{\partial_w^3 J^{-1}}{\partial_w J^{-1}} - \frac{3}{2} \Big( \frac{\partial^2_w J^{-1}}{\partial_w J^{-1}}\Big) = S[J^{-1}](w)\,,
\ee
with $S[f](w)$ the Schwarzian derivative of $f(w)$,
\be
S[f] = \frac{f'''}{f'} - \frac{3}{2}\Big(\frac{f''}{f'}\Big)^2\,.
\ee
We therefore find that in the metric \eqref{eq:metricons2}, the one-point function of the energy-momentum tensor is given by \eqref{eq:Tons2}. This is already an encouraging result: we mentioned above that the bulk geometry is captured by $J$ and here we find that the same $J$ arises in the boundary energy-momentum tensor, which therefore provides the holographic encoding of the bulk geometry. However, the boundary metric \eqref{eq:metricons2} also depends on $J$ which is not completely intuitive. This can be avoided by using $J$ once more to pull back everything to $H$. If we use a complex coordinate $z$ on $H$, so $w = J(z)$, then we find that:
\be
\label{eq:vevT}
\vev{T_{zz}} = - S[J] \qquad \qquad ds^2_{(0)} = \frac{dz d\bar z}{\text{Im}(z)^2}\,,
\ee
where we used that $(S[J^{-1}] \circ J)(dJ/dz)^{2} = - S[J]$, which follows from \cite{nag}
\be
\label{eq:transfschwarzian}
S[f \circ g] = (S[f]\circ g)(dg/dz)^2 + S[g].
\ee
This equation may be directly verified by using the chain rule for differentiation, which in our notation is written as $(f \circ g)' = (f' \circ g) g'$.

Equation \eqref{eq:vevT} is the result we are after: if we describe the boundary $S_d$ of the handlebody as the quotient $H/\Gamma_d$ (corresponding to the bottom picture in figure \ref{fig:J}), then the one-point function of the stress energy tensor in the constant negative curvature metric is given by minus the Schwarzian derivative of the map $J$ to $S^2$. If we now recall that $J$ dictates which cycles in $\Gamma_d$ are contractible in the bulk, namely precisely those for which $J \circ \gamma_d = J$, then this implies that $\vev{T_{zz}}$ indeed encodes the precise filling and therefore the bulk geometry.

Notice also that $S[J]$ has the right transformation
properties under composition of $J$ with $SL(2,\Real)$ from the right,
under which it transforms covariantly, and with $SL(2,\Complex)$ from
the left, under which it is invariant. These transformation properties
follow from \eqref{eq:transfschwarzian} and the fact that $S[f] = 0$
if $f$ is a M\"obius transformation \cite{nag}.

Finally, let us mention that the
renormalized on-shell bulk gravity action has been computed
in \cite{Krasnov:2000zq} and shown to be equal to the on-shell
Liouville action on $S_d$, computed earlier in the mathematics
literature \cite{TZ}. Note also that the map $J$ implicitly defines
a solution to the Liouville equation.

\subsection{Bulk interpretation and relation to Teichm\"uller theory}

We can make the holographic encoding of the spacetime a little more
explicit. As mentioned above, from the bulk perspective
the boundary Weyl rescaling is induced by a bulk diffeomorphism
that preserves the Fefferman-Graham form of the metric
but introduces a new Fefferman-Graham radial coordinate $\rho'$
\cite{Imbimbo:1999bj,Skenderis:2000in}. For the case at hand, the precise bulk
diffeomorphism is given in \cite{Krasnov:2001cu,mythesis} and leads to
the bulk metric
\be
\label{eq:metrichandlebody}
ds^2 = \frac{d\rho^2}{\rho^2} + \frac{(1+ \frac{1}{4}\rho^2)^2}{\rho^2} \frac{|dz + \mu_\rho d\bar z|^2}{\Im(z)^2}\,,
\ee
with $z$ again a coordinate on $H$ and
\be
\label{eq:muhandlebody}
\mu_\rho(z,\bar z) = - \frac{1}{2}\frac{\rho^2}{1 + \frac{1}{4}\rho^2}
\left(\overline{S[J](z)}\right)\Im(z)^2\,,
\ee
where the bar indicates complex conjugation and
we dropped the primes on the new coordinates. Indeed, by
expanding this metric in $\rho^2$ and using \eqref{eq:vevTgeneral} we
obtain again the result \eqref{eq:vevT}. It is noteworthy to mention
that in the new coordinates the action of $\Gamma$ leaves slices of
constant $\rho$ invariant, so its elements $\gamma$ just act as
$(\rho,z) \to (\rho,\gamma(z))$ with $\gamma(z)$ given by
\eqref{eq:mobius}.

We expect these new coordinates to become ill-defined somewhere inside the handlebody since the contractible cycles shrink to zero length at a certain point. By inspection of \eqref{eq:metrichandlebody}, this only happens when $|S[J](z)| \, \Im(z)^2 > \frac{1}{2}$. This bound on the Schwarzian derivative is familiar from Teichm\"uller theory as it figures prominently in the Ahlfors-Weil theorem concerning a local inverse of Bers' embedding of Teichm\"uller spaces \cite{nag} in the space of holomorphic quadratic differentials. The physical relevance of the bound is the following. When this bound is nowhere satisfied the coordinate system is nonsingular all the way to $\rho \to \infty$ where we recover another asymptotically AdS region. We then do not describe a wormhole but rather a spacetime with two disconnected boundaries which are simultaneously uniformized in the boundary $S^2$, as expected from Teichm\"uller theory. These do not correspond to wormholes and we refer to \cite{Maldacena:2004rf} for more information as well as open questions regarding these spaces. For a handlebody there are no other asymptotic regions and we may therefore assume on physical grounds that the bound is everywhere satisfied. In that case, the coordinate system becomes degenerate at a surface given by
\be
\rho^2 = \rho_c^2 \equiv \frac{1}{|S[J]|\, \Im(z)^2 - \frac{1}{2}}
\ee
At the surface $\rho = \rho_c$, the metric is everywhere degenerate since $|\mu_{\rho_c}| = 1$. We then describe a point in the boundary of Teichm\"uller's compactification of the Teichm\"uller space \cite{nag}. It would be interesting to verify explicitly that the contractible cycles are indeed the degenerate cycles on this surface.

\subsection{Non-handlebodies}

The discussion so far was about Euclidean handlebodies, but
these are not the only 3-manifolds that have
a genus $g$ Riemann surface as their conformal boundary.
We briefly discuss an example of such non-handlebody spacetimes in this
subsection\footnote{We thank Alex Maloney for discussions about
the material in this subsection.}.
A simple example can be constructed from the spacetimes described in \cite{Maldacena:2004rf}.
These are obtained by starting from $H^3$ written in hyperbolic slicing and
quotienting the boundary by a discrete subgroup $\G$ of $H$
to obtain a compact, finite volume, genus $g>1$ surface, $\Sigma_g$.
This yields the metric with two boundaries,
\be
ds^2 = dr^2 + \cosh^2 (r) ds^2_{\Sigma_g}
\ee
where $r \in (-\infty, \infty)$ and
\be
ds^2_{\Sigma_g} = \frac{dz d \bar z}{\Im(z)^2}
\ee
is the constant negative curvature metric on $\Sigma_g$ which
has scalar curvature $R = -2$. To produce a manifold
with a single boundary, one may try to quotient by
$r \to -r$. This procedure however introduces a singularity
at $r=0$. The singularity can be avoided if the
surface $\Sigma_g$ has a fixed point free involution $I$,
since then we can combine  $r \to -r$
together with the action of $I$ to obtain a smooth
hyperbolic 3-manifold with conformal boundary the Riemann
surface $\Sigma_g$. Such involutions are discussed, for
example, in \cite{Parlier}.
In this case the singularity at $r=0$
is replaced by the smooth Riemann surface $\Sigma_g/I$.
The resulting 3-manifold is a quotient of $H^3$ which has no
contractible cycles so it is not a handlebody.

This 3-manifold has the same conformal boundary as the
handlebody build from $\Sigma_d$ but it has a different
expectation value of the energy momentum tensor. Changing
variable, $\rho=2 e^{-r}$, the metric becomes of the form
(\ref{eq:barefgmetric}) with:
\be
g_{(0)ij} = 2 g_{(2)ij} = ds^2_{\Sigma_g}.
\ee
We may then use \eqref{eq:vevTgeneral} to obtain that:
\be
\vev{T_{ij}} = - g_{(0)ij}.
\ee
We see that the one-point function of the energy-momentum tensor is
notably different from that of a handlebody. However, we also observe
that any involution that `ends' the spacetime at $r=0$ (with or
without fixed points, orientation-reversing or orientation-preserving)
results in the same one-point function, so the holographic one-point
function of $T_{ij}$ does not seem able to distinguish these
geometries.

This is an interesting subtlety of the Euclidean dictionary
due to global issues.
Let us recall why we expect that locally,
in the Euclidean setup,  $g_{(0)ij}$ and $\vev{T_{ij}}$ uniquely fix
a bulk solution (in any dimension).
Intuitively, this is because the bulk equations of motion
are second order differential equations and  $g_{(0)ij}$ and $\vev{T_{ij}}$
provide the correct initial data. One
can indeed show rigorously that given this data
there exists a unique bulk solution in a thickening of the
conformal boundary, see \cite{Anderson:2004yi} and references therein.
Furthermore, one can show that $(g_{(0)ij}, T_{ij})$ are coordinates
in the covariant phase space of the theory \cite{Papadimitriou:2005ii}
and thus each such pair specifies a solution.
In the case at hand this data indeed produces a unique metric
for $r>0$ but the way the spacetime is capped off at $r=0$ depends on
the fixed point free involution used. One can presumably distinguish the
different spacetimes by using higher point functions and
non-local observables, such as the expectation
values of Wilson loops, \ie minimal surfaces that end
at a loop in the conformal boundary of the 3-manifold.
It would be interesting to verify this explicitly.

\section{Holographic interpretation of Lorentzian wormholes}
\label{sec:Lorentzian}

We now move to discuss the holographic interpretation of the
Lorentzian wormholes. We start by demonstrating in the next subsection that
a naive adaptation of the analysis of the previous section leads to
incomplete results where, in contrast with the Euclidean results, the spacetime geometry does not seem to be captured by the dual field theory on the Lorentzian side.
This is then resolved using the real-time
gauge/gravity prescription of \cite{us,us2}.

\subsection{Naive computation}

As we mentioned earlier, the metric in the outer regions can always be cast
in the BTZ form \eqref{eq:btzmetric}.
When using a new coordinate $\rho$ defined via
\be
r = \frac{\frac{M}{4}\rho^2 + 1}{\rho}\,,
\ee
the metric takes the form in (\ref{eq:barefgmetric})
with
\be
\label{eq:fgcoefficients}
g_{(0)ij} = \eta_{ij}\,, \qquad g_{(2)ij} = \frac{M}{2} \delta_{ij}\,, \qquad g_{(4)ij} = \frac{M^2}{16} \eta_{ij}\,.
\ee
The one-point function of the stress energy tensor in the dual state
can be computed from (\ref{eq:vevTgeneral}) yielding,
\be \label{eq:naive1pt}
\vev{T_{ij}} = M \delta_{ij}\,.
\ee
On the other $(m-1)$ conformal boundaries, we obtain similar one-point
functions (with different values of $M$) and all the other one-point
functions vanish. This is problematic, since we obtain no information
whatsoever about the inner part of the geometry and the Lorentzian
one-point functions therefore seem to be insufficient to reconstruct
the wormhole spacetime. The holographic encoding of the spacetime
appears to fail, which would contradict standard expectations from the
gauge/gravity duality. This apparent contradiction comes from the
fact that we have not taken into account the holographic
interpretation of Cauchy data. This can done using the
real-time gauge/gravity prescription of \cite{us,us2}, which we
review in the next subsection.

\subsection{Lorentzian gauge/gravity prescription}

In this subsection, we prepare for the discussion below by reviewing
some known facts about Lorentzian quantum field theory. Afterwards, we show how
one may use these facts to obtain a consistent real-time prescription
for the gauge/gravity duality. We then return to the wormholes in
subsection \ref{sec:wormholestates}.

\subsubsection{States in field theory}
\begin{wrapfigure}{R}{4cm}
\centering
\psfrag{C1}{$C_1$}
\psfrag{C2}{$C_2$}
\includegraphics[width=3cm]{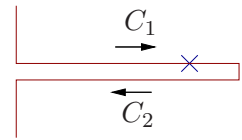}
\caption{\label{fig:simplecontour}A contour in the complex time plane; the cross signifies the operator insertion.}
\end{wrapfigure}

The prescription in \cite{us,us2} is based on the fact that \emph{any}
Lorentzian field theory path integral requires a specification of the
initial and final states as well. Such a state $\ket{\Psi}$
may be specified via path integrals on a Euclidean space $Y$ with a
boundary and possible operator insertions away from this boundary. If
we want to compute, say, $\vev{\Psi|\op(t)|\Psi} = \vev{\Psi|e^{i H
    t}\op e^{-iHt}|\Psi}$, we continue to path integrate along a
Lorentzian segment with length $t$ that is glued to the boundary of
the Euclidean space, then insert the operator, and finally go
\emph{back} in time for a period $t$ before we attach a second copy of
the Euclidean space. For Euclidean spaces which are topologically
$\Real \times X$ with $X$ a real space and $\Real$ representing
Euclidean time, the overall field theory background manifold
corresponds to a contour in the complex boundary time plane of the
form sketched in figure \ref{fig:simplecontour} times a real
space. Notice that extending the contour beyond the point $t$, say to
a point $T > t$, amounts to an extra insertion of $e^{iH(T-t)}
e^{-iH(T-t)} = \mathbf 1$ which does not affect the correlation
function. A similar story holds for higher-point correlation
functions, but in those cases an operator ordering has to be
specified. Although the contour may often be deformed to a simpler
version, we emphasize that a procedure like the above is \emph{always}
necessary for Lorentzian quantum field theory.

\subsubsection{Translation to gravity}

In the Lorentzian gauge/gravity prescription of \cite{us,us2}, one
incorporates the Euclidean segments for the path integral into the
holographic description and `fills' them with a bulk solution as
well. For example, to the contour of figure \ref{fig:simplecontour}
may correspond a bulk manifold consisting of two Lorentzian and two
Euclidean segments. These segments are then glued to each other along
spacelike hypersurfaces that should end on the corners of the boundary
contour. The behavior of the fields at these hypersurfaces is then
determined using \emph{matching conditions}.
These guarantee the $C^1$ continuity of the fields. More
precisely, for the metric one imposes continuity of the induced metric
$h_{AB}$ and the extrinsic curvature $K_{AB}$ with a factor of $i$:
\be \label{matching}
^L\! h_{AB} = {}^E\! h_{AB},\qquad \qquad ^L\! K_{AB} = -i \ ^E\! K_{AB},
\ee
with the superscript indicating the Lorentzian or the Euclidean side
and the extrinsic curvature on either side is defined using the
outward pointing unit normal. There is also a \emph{corner} matching
condition\footnote{It is likely that this condition
follows from the matching for the induced metric
and the extrinsic curvature in (\ref{matching}), but
in the absence of a general proof we
treat it as an additional matching condition.},
which is defined at the intersection between $S$ and the
conformal (radial) boundary. It dictates that the inner product
between the unit normal to $S$, denoted as $n_\mu$, and the unit
normal to the radial boundary, written as $\hat n_\mu$, is continuous
across the boundary (up to appropriate factors of $i$). For a
Lorentzian-Euclidean gluing, using outward pointing unit normals, it
becomes:
\be
^L\! (\hat n^\mu n_\mu) = i \ {}^E\! (\hat n^\mu n_\mu).
\ee
 As discussed in \cite{us2}, all the matching conditions arise naturally from a saddle-point approximation. Although they are equivalent to analytic continuation in many simple cases, they do not rely on a globally defined time coordinate and are therefore more generally applicable.

This construction is an essential ingredient in the Lorentzian
gauge/gravity dictionary. For example, it allows us to understand
precisely how changing the initial and final states modifies the
Lorentzian spacetime, gives the correct initial and final conditions
for the bulk-boundary and bulk-bulk propagators, and also cancels
surface terms from timelike infinity in the on-shell action, which
would otherwise lead to additional infinities.
Furthermore, the boundary correlators directly come in the
in-in form as in quantum field theory.

\subsection{Gauge/gravity duality for Lorentzian wormholes}
\label{sec:wormholestates}

Let us now apply the construction outlined in the previous
subsection to the wormholes. To this end, we have to cut off the
wormhole along some spatial bulk hypersurface and find a Euclidean
space that we may glue to this hypersurface such that the matching
conditions are satisfied.

Of course the field theory contour also has a backward-going segment
and  a final state. To fill this in, we have to cut off the wormhole
along some final time slice as well and glue a second Lorentzian and
Euclidean segment to this final surface. These second copies can be
taken to be identical to the first ones, which correspond to taking the
final and the initial state to be just the same. In \cite{us2}, we
performed this procedure for the eternal BTZ black hole. As long as we
do not switch on any perturbations, we may take the second Lorentzian
and Euclidean segment to be completely identical to the first
one. This also means that the matching conditions are trivially
satisfied along the final gluing surface, so these do not have to be
investigated separately. Therefore, it will be sufficient to focus on
a single Euclidean-Lorentzian gluing below.

\begin{figure}
\centering
\includegraphics[width=10cm]{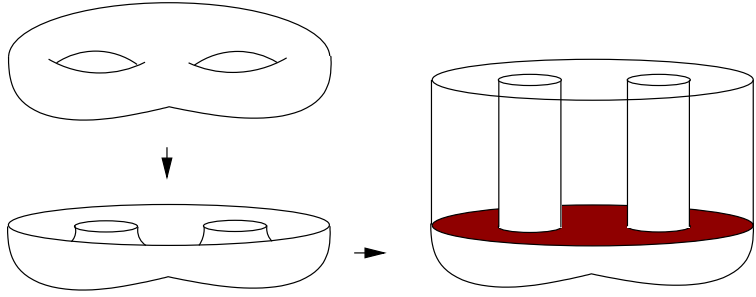}
\caption{\label{fig:gluedwormhole0}We take half of a genus two surface and attach Lorentzian cylinders to the boundary. This boundary manifold can be filled in with half an Euclidean handlebody plus a Lorentzian wormhole with spatial topology of a pair of pants. On the right, we shaded the matching surface between the Euclidean and the Lorentzian segment. It indeed has three boundaries and no handles.}
\end{figure}
\begin{figure}
\centering
\includegraphics[width=10cm]{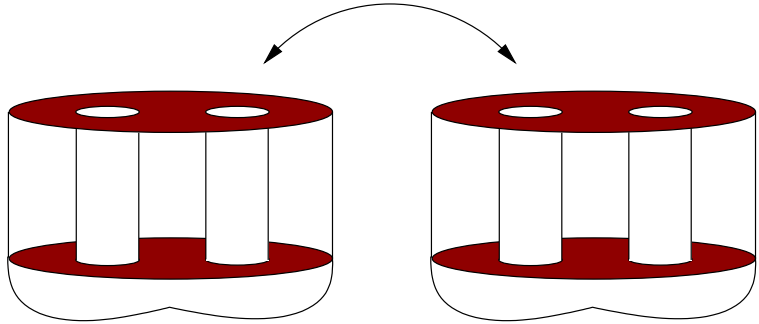}
\caption{\label{fig:gluedwormhole5}Analogous to figure \ref{fig:simplecontour}, the full field theory contour has a forward- and a backward-going segment and two Euclidean segments to specify the initial and final state. Similarly, the full bulk spacetime consists of four segments as well. They should be glued along the matching surfaces which we shaded in this picture.}
\end{figure}

A candidate for the Euclidean space is half of the Euclidean
handlebody $M_e$ that we obtained in section
\ref{sec:euclidean}. (It is not however the \emph{only}
candidate, as we will explicitly demonstrate in subsection
\ref{sec:morefillings}.)
Indeed, we may cut this handlebody and the wormhole spacetime in two
halves along the surface $S$ and glue them together along $S$. In the
case $S$ is a pair of pants (a surface of genus zero with three
circular boundaries, so $g=0$ and $m=3$), the procedure is sketched in
figure \ref{fig:gluedwormhole0} and the filling of the full field
theory contour, including the backward-going segment, is sketched in
figure \ref{fig:gluedwormhole5}. Let us now verify that the matching
conditions are satisfied at the shaded matching surface in figure
\ref{fig:gluedwormhole0}. On both sides, the induced metric is locally
just the unique negative curvature metric on $S$ described as
$H/\Gamma$, so it is the same metric indeed. Also, the extrinsic
curvature vanishes completely on both sides because of the $\mathbb
Z_2$ time-reversal symmetry. Therefore, the first and second matching
conditions are satisfied indeed. Finally, the extra corner matching
discussed in \cite{us2} is also satisfied: in our case $S$ intersects
the conformal boundary orthogonally (again because of the $\mathbb
Z_2$ symmetry) and the inner product $n^\mu \hat n_\mu$ thus vanishes
both for $M$ and for $M_e$. However, there is still a subtlety with
the boundary metric which we now discuss.

If we use the BTZ coordinate system on the Lorentzian side, then the
boundary metric on this side is flat. The boundary metric on the
Euclidean side, however, can never be globally flat because $S_d$ has
negative Euler number. On the other hand, to match $M$ and $M_e$, we
should also take the boundary metric to be smooth (in the sense
specified in \cite{us2}). This can be done by Weyl rescaling
the Lorentzian boundary metric to
a metric of constant negative curvature, as we discuss below.
The boundary metric is then smooth
across the corner and the discrepancy between the boundary metrics
on either side is removed.\footnote{Another possibility would be to Weyl rescale the metric on the Euclidean side such that it is flat in the vicinity of the gluing circles. Although the gluing is then smooth, the Euclidean boundary metric can then no longer be analytic.}

\subsubsection{Matching Euclidean and Lorentzian  wormholes and $\vev{T_{ij}}$ }

\begin{figure}
\centering
\psfrag{Lor}{}
\psfrag{t}{$\tau$}
\psfrag{x}{$x$}
\psfrag{l}{$l$}
\psfrag{w}{$w$}
\includegraphics[width=5cm]{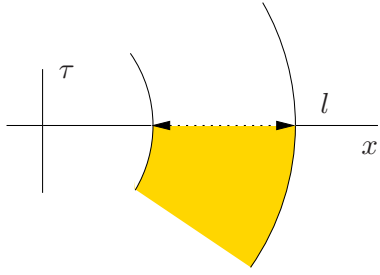}
\caption{\label{fig:muz}Part of a fundamental domain of $S_d$ on $S^2$. We will eventually replace the part with $\tau > 0$ with a Lorentzian wormhole. The single identification is given by $w \sim \lambda w$ and the line $l$ is the entire positive $x$-axis.}
\end{figure}

We now discuss the consequences of the continuity of the boundary
metric across the matching surface. As described above we match
the initial $U=0$ surface of the Lorentzian wormhole to half of the
Euclidean handlebody. On the boundary of the spacetime, the Lorentzian
cylinders are glued to the boundary of the Euclidean handlebody along
the $m$ circles that form the boundary of the $U=0$ Riemann
surface. These $m$ circles lift to segments of the great circle given
by $\tau = 0$ in the Poincar\'e coordinates \eqref{eq:euclpoinc} on
the boundary $S^2$ of $H^3$. Let us now focus on one of the
$m$ circles. After conjugation, we can always ensure that it lifts
to the half-line $l$ given by:
\be \label{eq:l_line}
l: x > 0, \tau = 0
\ee
on the $S^2$. Its projection down to $S_d$ is then given via the identification
\be
\label{eq:xtlambda}
w \sim \lambda w
\ee
with $w = x+ i\tau$ and for some positive real $\lambda \neq 1$.
The relevant part of the fundamental domain is then sketched in
figure \ref{fig:muz}.

\begin{figure}
\centering
\psfrag{Lor}{}
\psfrag{z}{$z$}
\psfrag{l}{$l'$}
\includegraphics[width=4cm]{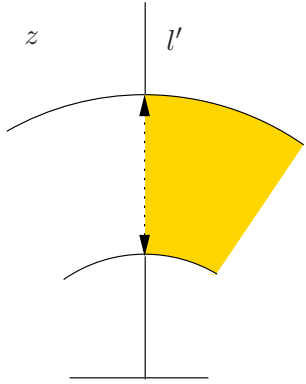}
\caption{\label{fig:lambdaz}Under the locally defined map $J^{-1}$ the domain in figure \ref{fig:muz} maps to the sketched domain in $H$, where we use a coordinate $z$. The identification is given by $z \sim \mu z$. The line $l'$ is the positive imaginary axis. We will replace the part $\Re(z) < 0$ with a Lorentzian wormhole.}
\end{figure}

To find the boundary metric of constant curvature on this surface we again have to pass from the description of $S_d$ as quotient of
$(S^2 \backslash \Lambda(\hat \Gamma))$  to that of a quotient of $H$, for which we defined the map $J$ in section \ref{sec:holographic}.
Using $J^{-1}$, we now map
the half-line (\ref{eq:l_line}) to $H$, where we use the
coordinate $z$. Although $J^{-1}$ is multi-valued, we will need only
one of the images of $l$ in $H$. We can again use conjugation freedom
to make sure that the image under consideration is the half-line:
\be
l': \Re(z) = 0\,.
\ee
In $H$, the identification \eqref{eq:xtlambda} becomes an isometry of
$SL(2,\Real)$ that leaves $l'$ invariant. Such an isometry is
necessarily of the form:
\be
z \sim \mu z\,,
\ee
for some real $\mu \neq 1$ given implicitly by
\be
J(\mu z) = \lambda J(z).
\ee
The construction in $H$ is sketched in figure \ref{fig:lambdaz}.
Notice that $J(z)$ is an analytic map from the imaginary axis to the
(positive) real axis, that is
\be
\overline{J(iy)} = J(iy), \qquad  y >0.
\ee
Notice also that the $\mathbb Z_2$ symmetry $w
\leftrightarrow \bar w$ maps under $J^{-1}$ to reflection in the
imaginary axis, that is $z \leftrightarrow -\bar z$. (Again, as
$J^{-1}$ is multi-valued, it maps the original $\mathbb Z_2$ to many
other reflections in $H$ as well, but we do not need them here.)

\begin{figure}
\centering
\psfrag{zp}{$z'$}
\psfrag{u}{$u$}
\psfrag{v}{$v$}
\includegraphics[width=11cm]{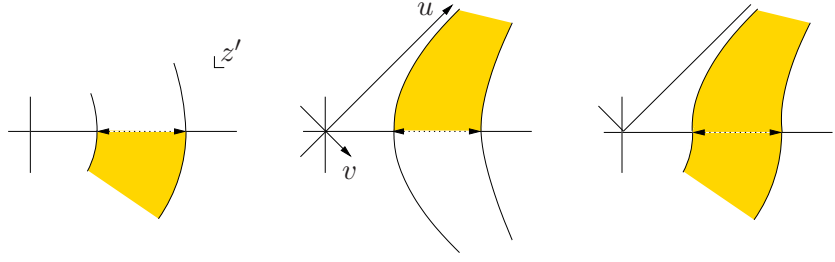}
\caption{\label{fig:gluing}On the left, the Euclidean boundary geometry in the coordinate $z'$. In the center figure we sketched the Lorentzian boundary geometry and on the right the glued-together geometry.}
\end{figure}

We now ready to attach a Lorentzian cylinder to the boundary.
The procedure is sketched in figure \ref{fig:gluing}. On $H$,
this means that we cut away the half given by $\Re(z) < 0$
and attach the universal covering of a Lorentzian cylinder to the
gluing line $\Re(z) = 0$. In the bulk, we can use the metric
\eqref{eq:metrichandlebody} with the matching surface given by $\Re(z)
= 0$, at least up to the point $\rho = \rho_c$. We now need to find a
Lorentzian bulk metric that satisfies the matching conditions of
\cite{us2} when glued to this surface.

Both in the bulk and on the boundary, it is straightforward to obtain
the explicit matching Lorentzian metric by analytic continuation. We
first introduce a coordinate $z' = -iz$. In the $z'$ plane the figure
\ref{fig:lambdaz} is rotated clockwise by 90 degrees which slightly
simplifies the matching below. In the coordinate $z'$ the metric
\eqref{eq:metrichandlebody} becomes:
\be
ds^2 = \frac{d\rho^2}{\rho^2} + \frac{(1+ \frac{1}{4}\rho^2)^2}{\rho^2} \frac{|dz' + \mu_\rho d\bar z'|^2}{\Re(z')^2}\,,
\ee
where now
\be
\mu_\rho(z',\bar z') = \frac{1}{2}\frac{\rho^2}{1 + \frac{1}{4}\rho^2}
\left(\overline{S[\tilde{J}](z')}\right)\Re(z')^2\,,
\qquad \tilde{J}(z')= J(i z')
\ee
where we used that $S[J](iz') = - S[\tilde{J}](z')$,
which follows from \eqref{eq:transfschwarzian}. The gluing takes place along the half-line $\Im(z') = 0$, $\Re(z') > 0$. We then replace $z' \to u$ and $\bar z' \to v$ to find the Lorentzian bulk metric:
\be
\label{eq:lormetrichandlebody}
ds^2 = \frac{d\rho^2}{\rho^2} + \frac{(1+ \frac{1}{4}\rho^2)^2}{\rho^2} \frac{(du + \mu_\rho (v) d v)(dv + \mu_\rho (u) du)}{\frac{1}{4}(u + v)^2}\,,
\ee
with
\be
\mu_\rho(u) = \frac{1}{8}\frac{\rho^2}{1 + \frac{1}{4}\rho^2}
\Big(S[\tilde{J}](u)\Big)\, (u+v)^2\,,
\ee
and a similar expression with $u \to v$. Note that
$\tilde{J}(x)$ is real-analytic for $x>0$ and monotonic, so
$S[\tilde{J}](x)$ is real-analytic too.
This Lorentzian metric is thus real and covers the bulk spacetime up to $\rho = \rho_c$. Since $z' \sim \mu z'$, the periodicity on the Lorentzian side is $(u,v) \sim \mu (u,v)$. The point $(u,v) = (0,0)$ on the boundary is a fixed point of this identification and therefore we need to exclude the forward lightcone emanating from this point from the spacetime (the backward lightcone is already replaced by the Euclidean geometry). Since we also demanded $\Re(z') > 0$, so $u + v > 0$, we need only the part of the Lorentzian boundary with $u>0$ and $v>0$.

On the boundary we find the metric:
\be
ds_{(0)}^2 = \frac{du dv}{\frac{1}{4}(u+v)^2}
\ee
which has scalar curvature $R_{(0)} = -2$. Using once more \eqref{eq:vevTgeneral} we obtain for the one-point functions:
\be \label{eq:1pt}
\vev{T_{uu}} = - S[\tilde{J}](u) \qquad \qquad
\vev{T_{vv}} = - S[\tilde{J}](v) \qquad \qquad \vev{T_{uv}} = \frac{-1}{8(u+v)^2}
\ee
Notice that one expects that $T_i^i = \frac{c}{24 \pi} R_{(0)}$ and
we obtained here  $T_i^i = -2$.  Reinstating the factors of
$16 \pi G_N $, we find $c = 24\pi/(16 \pi G_N) =
3/(2G_N)$ which is indeed the correct central charge.

Equation \eqref{eq:1pt} is the main result of this section and
demonstrates that the Lorentzian one-point function of the stress energy
tensor as obtained from the metric \eqref{eq:lormetrichandlebody} does
contain information about the dual geometry that is hidden behind the
horizons.

\subsection{More fillings}
\label{sec:morefillings}

In the previous section we glued a particular handlebody to the Lorentzian
wormhole. There exist a variety of handlebodies $\{M_e\}$ that all have
a hypersurface $S$ where the matching conditions are satisfied as we discuss
now.

\begin{wrapfigure}{L}{5cm}
\centering
\includegraphics[width=5cm]{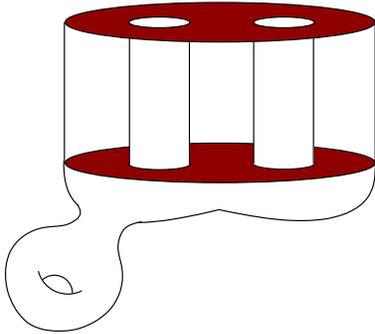}
\caption{\label{fig:gluedwormhole4}Adding a handle as indicated does not change the properties of the gluing surface or the Lorentzian spacetime.}
\end{wrapfigure}

\begin{figure}[b]
\centering
\psfrag{t = 0}{$\tau = 0$}
\includegraphics[width=12cm]{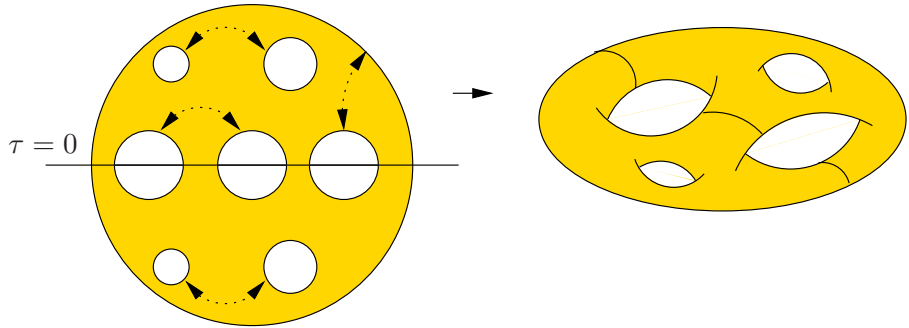}
\caption{\label{fig:schottkyfundamentalissues}As indicated on the left, one may add generators to the Schottky group $\hat \Gamma$ without breaking the $\mathbb Z_2$ symmetry. The resulting surface has two extra handles: one for the half corresponding to the initial state and another one for the final state.}
\end{figure}

In particular, one may attach an
extra filled handle to $M_e$ somewhere away from the matching surface
to obtain a manifold $M'_e$ with conformal boundary $\Sigma$ with
$\Sigma \neq S_d$. An example of this is sketched in figure
\ref{fig:gluedwormhole4}. This
procedure does not change any properties like the induced metric or
the extrinsic curvature of the matching surface. Geometrically, this
can be seen by going to the universal covering: one may add generators
to $\hat \Gamma$ to obtain a group $\hat \Gamma'$ and as long as $M_e'
= H^3/\hat \Gamma'$ has $S$ as a surface of $\mathbb Z_2$ symmetry we
may slice open $M_e'$ along this surface and glue the Lorentzian
wormhole $M_l = \wAdst/\hat \Gamma$ to it. For the boundary surface
this procedure is sketched in figure
\ref{fig:schottkyfundamentalissues}. In the figure we represented the
addition of two generators to $\hat \Gamma$ by cutting out four
circles out of the fundamental domain that are pairwise identified,
all done in such a way that the original $\mathbb Z_2$ symmetry
remains intact. Although we have not sketched it here, this procedure
directly extends to the entire three-dimensional space: the $\mathbb
Z_2$ symmetry is also present for the new handlebody and $S$ is again
the invariant surface given by $\tau = 0$.

We conclude that we can glue to
the Lorentzian wormholes also half of $M_e'$. A similar
analysis as in the previous section establishes that the 1-point functions
captures the fact that the initial state is different than the one
corresponding to $M_e$.

\subsection{State dual to wormholes}

Let us now discuss what our results imply about the QFT state dual
to the wormholes. Since there are $m$ boundaries, the Hilbert
space consists of a tensor product of $m$ Hilbert spaces, one
for each boundary component. From the fact that the wormholes
are manifolds that interpolate between the $m$ segments,
we expect to find nonzero correlations between the $m$ boundaries
and the initial state to be an entangled state.
Indeed, this is precisely what we find. To see this,
suppose the initial state is separable, namely of a product form
$|\a_1\rangle \otimes \cdots \otimes |\a_m\rangle$.
Then the 1-point functions would necessarily take a factorizable form.
More precisely, suppose the state was separable and
consider the insertion of a stress energy tensor
in, say, the first boundary component,
\bea
\langle T_{ij}(x_1)\rangle &=&
\langle \a_1| \otimes \cdots \otimes \langle \a_m|  T_{ij}(x_1)
|\a_1\rangle \otimes \cdots \otimes |\a_m\rangle \nonumber \\
&=& \langle \a_1| T_{ij}(x_1) |\a_1\rangle \prod_{k=2}^2
|| |\a_k\rangle ||^2
\eea
Now the naive 1-point function
in (\ref{eq:naive1pt}) would support the view that the state is
separable. Namely, in that case we could naively say that the
state $|\a_1\rangle$ in the first copy depends only on the
corresponding mass parameter $M_1$ and not on the other
variables that determine the spacetime. This would lead to
one- and higher-point functions of the energy momentum
tensor in the first copy which up to an overall factor only
depend on $M_1$. The one-point functions that we got,
however, in (\ref{eq:1pt})
are not of  that form, as the Schwarzian $S[\tilde{J}]$ does not
have such a factorizable form and does contain all the
variables that determine the spacetime.

Another check on the non-separability of the state is provided
by the computation of a two-point function. An argument analogous
to the one above implies that if the state is separable then
the 2-point function would have a factorizable form. We
illustrate that this is not the case in the next subsection.

A natural guess for the dual state is that it is the state
obtained by an Euclidean path integral over a Riemann surface
$\Sigma$ with $m$ circular boundaries. According to the reasoning of
\cite{Maldacena:2001kr,Freivogel:2005qh},
this surface $\Sigma$ can be taken to be
precisely the conformal boundary of one half of the Euclidean manifold
$M_e$. This can be  $\Sigma = S$, with $S$ the
surface of time reversal symmetry of the wormhole spacetime,
for the case of the handlebodies of section \ref{sec:euclidean},
or  $\Sigma \neq S$ if the initial state is that of
the previous subsection.

If we now trace over all components but one,
all wormholes with $m > 1$ can be thought of as
been associated with a mixed state in the remaining copy
and this explains the presence of horizons.
The $m=1$ case is special in that we only have a
single copy of the CFT so there are no copies to trace out.
Nevertheless results for the 1- and 2-point functions
indicate that there is an entanglement between the outer
region and the region behind the horizon.
These spacetimes were also analyzed in \cite{Louko:1998hc}
which suggested that the dual state is in some respects similar
to a thermal state. We leave a better understanding of this case
for future work.

\subsection{2-point functions}

We discuss in this subsection the computation of the 2-point function
for a scalar operator $\op$ of dimension $\Delta$.
In the bulk it suffices to consider a free massive scalar
field, as interaction terms contribute only to higher point functions.
We glue  an Euclidean handlebody at $t=0$ and take the initial and
final states to be the same. It follows from an analysis along the lines
of \cite{us2} that the different real-time correlators (time-ordered,
Wightman, etc.) are obtained by suitable analytic continuations of the
Euclidean correlator in the handlebody geometry.

The two-point function of a scalar operator on the Euclidean plane
is uniquely fixed by conformal
invariance and takes the form:
\be
\vev{\op(\tau,x)\op(\hat \tau,\hat x)} = \frac{1}{[(\tau - \hat \tau)^2 + (x - \hat x)^2]^{\Delta}},
\ee
where we normalized the operators so that the coefficient in
the numerator equals one.
For the handlebody, we have to sum over the elements of the Schottky group
$\hat \Gamma$, whose elements $\gamma$ act as
M\"obius transformations on the boundary,
\be
\gamma: \omega=x + i \tau \to \frac{a(x + i\tau) + b}{c (x + i\tau) + d}\,,
\ee
with real $a,b,c,d$ and $ad -bc = 1$. This can also be written as
\be
\gamma: \Big(\tau,x\Big) \to \Big(\gamma_\tau, \gamma_x\Big) \equiv \frac{1}{(cx + d)^2 + c^2 \tau^2}\Big(\tau, (a x + b) (c x + d) + ac \tau^2 \Big)\, .
\ee
Using the complex coordinate $w$ on the boundary $S^2$ of $H^3$, we obtain
\be
\label{eq:twopointeuclpoinc}
\begin{split}
\vev{\op(w,\bar w)\op(w_1, \bar w_1)} = \sum_{\gamma \in \hat \Gamma} \frac{1}{|c w + d|^{2\Delta} |\gamma_w - w_1|^{2\Delta}}\,.
\end{split}
\ee
where the boundary metric is locally $dw d\bar w$. We then Weyl transform to the metric \eqref{eq:metricons2} which is globally well-defined to find
\be
\vev{\op(w,\bar w)\op(w_1, \bar w_1)} = \sum_{\gamma \in \hat \Gamma} \frac{e^{-\Delta \sigma(w,\bar w)} e^{-\Delta \sigma(w_1,\bar w_1)}}{|c w + d|^{2\Delta} |\gamma_w - w_1|^{2\Delta}}\,, \qquad ds^2 = e^{2\sigma(w,\bar w)}dw d\bar w\,,
\ee
with
\be
e^{2\sigma(w,\bar w)} = \Big|\frac{dJ^{-1}}{dw}\Big|^2 \frac{1}{\text{Im}(J^{-1}(w))^2} \,.
\ee
We can now pull back to $H$, using $z = J^{-1}(w)$, to obtain
\be
\vev{\op(z,\bar z)\op(z_1, \bar z_1)} = \sum_{\gamma \in \hat \Gamma} \frac{| J'(z) J'(z_1)|^\Delta (\Im(z)\Im(z_1))^\Delta }{|c J(z) + d|^{2\Delta}|\gamma(J(z)) - J(z_1)|^{2\Delta}}\,, \qquad ds^2 = \frac{dz d\bar z}{\Im^2(z)}\,.
\ee
As before, we may assume the covering groups are such that $J(\lambda z) = \mu J(z)$. We then again introduce the coordinate $z' = -iz$ and the map $\tilde J(z') = J (iz') = J(z)$, replace $z' \to u$ and $\bar z' \to v$ to obtain the Lorentzian metric. We recall that in this case $\tilde J(x)$ is real-analytic for real positive $x$. More precisely, following the steps in \cite{us2}, one finds
that the time-ordered correlator is obtained by
replacing $z \to u -i \epsilon u$ and
$\bar z \to v + i \epsilon v$, where the $i\epsilon$
insertions push the singularity everywhere away from the real-time
contour.
To avoid clutter we will however not write the $i\epsilon$
insertions explicitly below. The final answer is then
\be
\label{eq:twopointfunctionuv}
\vev{T \op(u,v)\op(u_1, v_1)} = \sum_{\gamma \in \hat \Gamma} \frac{2^{-2\Delta} (u+v)^\Delta (u_1 + v_1)^\Delta [\tilde J'(u) \tilde J'(v) \tilde J'(u_1) \tilde J'(v_1)]^{\Delta/2}}{(c \tilde J(u) + d)^\Delta(c \tilde J(v) + d)^\Delta(\gamma(\tilde J(u)) - \tilde J(u_1))^{\Delta}(\gamma(\tilde J(v)) - \tilde J(v_1))^{\Delta}}
\ee
in the metric
\be
ds^2 = \frac{4 du dv}{(u + v)^2}.
\ee
This 2-point function does not take a factorizable form supporting
the view that the dual state is entangled, as anticipated.
We would like to note however that at late times the correlator
is dominated by the BTZ elements in $\hat{\G}$. More precisely,
if one defines coordinates\footnote
{In these coordinates the  Lorentzian cylinder
is $(t, x) \sim (t, x + \log \l)$.} $u = \exp (x+t)$, $v = \exp (x-t)$,
then in the limit $t, t_1 \to \infty$ with $(t-t_1)$ fixed, all terms in
(\ref{eq:twopointfunctionuv}) go to zero, except the ones
with either $b=0$ or $c=0$. These are precisely the elements
associated with the BTZ black hole.

\section{Remarks}
\label{sec:remarks}
In this section we discuss some general remarks concerning the wormhole spacetimes.

\subsection{Other bulk spacetimes}
The question we addressed in this paper is what is the holographic
interpretation of any given wormhole spacetime. One can also ask:
given a geometry at infinity, how many different bulk spacetimes
can one have? In general, all such saddle points contribute
and should be taken into account, although typically
one of the saddle points dominates at large $N$ at any given
regime. A well-known example is that associated with the
Hawking-Page transition \cite{Hawking:1982dh,Witten:1998qj}.
In the case the boundary is
$S^1 \times S^{d-1}$ and there are two possible
(Euclidean) bulk manifolds corresponding to
making contractible in the interior either $S^1$ or $S^{d-1}$, namely
the Euclidean Schwarzschild AdS solution and thermal AdS.
This question is usually addressed in Euclidean
signature, but it is clearly also relevant in Lorentzian signature.
In this context the question is now: given the conformal boundary
of the complete Euclidean and Lorentzian pieces
how many different bulk manifolds
can one have?

Naively, one might think that for every Euclidean solution
there would be a corresponding Lorentzian plus Euclidean
solution, but this turns out not to be the case. This can
be demonstrated  with the case where the
conformal boundary is a torus, $S^1 \times S^1$.
As in the higher dimensional case, there are two
solutions that correspond to either the first or the second circle
being contractible in the interior (which correspond to thermal AdS
and Euclidean BTZ),
but there are now new possibilities obtained by
considering a contractible cycle that is a
linear combination of the above cycles \cite{Maldacena:1998bw}.
These solutions are called the `$SL(2,\mathbb Z)$ family' of
black holes as they related to the Euclidean BTZ black hole
by a modular transformation. In appendix \ref{app:fillings} we however show
that none
of these solutions can be used in the real-time gauge/gravity
prescription as the Euclidean part associated
with a vertical segment of the QFT contour.
The reason is that the matching conditions force
the bulk Lorentzian bulk metric to be
complex and this results in a energy momentum tensor
that does not satisfy the correct reality conditions.

\begin{figure}
\centering
\includegraphics[width=5cm]{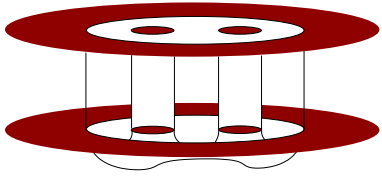}
\caption{\label{fig:thermal_m} The analogue of thermal AdS for the
pair of pants wormhole. In this case three copies of Lorentzian AdS$_3$
are attached to the three boundaries of the pair of pants.}
\end{figure}

For a higher genus Riemann surface there also exists a similar family of solutions
\cite{Yin:2007gv} as well as the aforementioned non-handlebody solutions.
As for the other fillings of handlebody-type, we expect that only the analogues of the
BTZ and the thermal AdS would be relevant for holography of the Lorentzian
wormholes. The analogue of thermal AdS is obtained by attaching
$m$ copies of empty Lorentzian AdS to the $m$ boundary components
of the handlebody. The case of a pair of pants wormhole is sketched
in figure \ref{fig:thermal_m}. For the non-handlebodies the corresponding
Lorentzian solution remains to be investigated.

\subsection{Rotating wormholes}

In the previous sections we considered non-rotating wormholes.
Rotating wormholes do exist \cite{Aminneborg:1998si,Brill:1999xm}
and are obtained by taking a quotient with respect to a group
generated by elements
of the form $(\gamma_1,\gamma_2) \in SL(2,\Real)\times SL(2,\Real)$
with $\gamma_1 \neq \gamma_2$. A similar region like $\wAdst$
exists such that the quotient $\wAdst/\Gamma$ is a good spacetime
\cite{Barbot:2005fv,Barbot:2005qk} and the metric in the
outer regions is isometric to the rotating BTZ metric
\cite{Barbot:2005qk}. The corresponding
`Euclidean spaces' for these wormholes, however, are not so straightforward.
A prescription for obtaining them has been proposed in
\cite{Krasnov:2001va} and was critically analyzed in \cite{mythesis}.
From the holographic perspective, the reality condition of the
bulk fields, especially on the Euclidean caps, should be dictated
by the standard reality condition of the dual QFT. In the case
of the rotating BTZ we have demonstrated in \cite{us2} (section 4.5)
that the matching conditions result in a complex metric on the Euclidean
caps. It is likely that the same would be true here, namely the
Euclidean solution that should be glued to the rotating Lorentzian
wormhole would be complex.

There are several issues that need to be resolved in order to understand the
rotating case. Firstly, it is not straightforward to find in the rotating wormhole
the analogue of a $U=0$ slice of the non-rotating wormhole \cite{Barbot:2005qk}.
One approach to this problem is to consider the rotating wormholes as deformations of the non-rotating wormholes. In the Lorentzian case such deformations might be described by a Lorentzian version of the standard quasiconformal mappings \cite{nag}, one for each $SL(2,\Real)$ factor. One would then need to extend these deformations to
the 'Euclidean' solutions, which, as mentioned above, are likely to be complex solutions that
possess a real slice where the Lorentzian solution can be glued. It would be interesting to further develop this direction.

\section{Outlook} \label{sec:conclusions}

We have discussed the holographic interpretation of a class of
2+1-dimensional wormhole spacetimes. They are interesting toy models
for the analysis of global issues in the real-time gauge/gravity
correspondence. We have shown that the asymptotics of the complete
solution that includes both the Lorentzian solution and the Euclidean
caps completely characterize the geometry including the regions
behind the horizons. This came about by a subtle interplay between
global issues and the real-time gauge/gravity dictionary.
In particular, the real-time gauge/gravity prescription requires
gluing smoothly Euclidean solution to the Lorentzian solution
at early and late times. This in turn fixes the apparent freedom for
independent Weyl rescaling at different outer components and
results in holographic data that contain information about the complete
geometry.

We thus find that the Lorentzian CFT correlators encode in a
very precise sense the parts of the geometry that lie behind
the horizons. This presents a unique opportunity to study
and settle classic questions and puzzles in black hole physics.
The way the information is given to us, however,
(i.e. in terms of CFT correlators) is very different
from the way the black holes puzzles are usually formulated (e.g.
using bulk local observers) and this presents the main
obstacle in directly addressing these issues.

In this respect, one of the most interesting cases to further
understand is that of spacetimes with $m=1$ and $g > 0$.
As discussed earlier, this has only one outer region. The form of the 1-point
and 2-point functions indicate entanglement between the outer
region and the region behind the horizon. It is not clear however
which modes are entangled in the CFT, since unlike the cases with $m > 1$ the dual state
seems to be defined in only a single copy of the Hilbert space.

We can however suggest some
possibilities. Note that all wormholes can be viewed as quotients of a part of
BTZ, since the group $\Gamma$ associated with them always
contains a subgroup isomorphic to that of BTZ (namely $\mathbb Z$)
and so one can take the quotient first with respect to this group,
resulting in BTZ, and then with respect to the rest of the group
elements (modulo issues related to the regions one needs to
remove to avoid closed timelike curves that need to be investigated).
Thus we find a state in the tensor product of two Hilbert spaces
(associated with the two boundaries of BTZ) with certain correlations
between the two components because of the final quotient.
It would be interesting to make this more precise and understand its
relation with the apparent entanglement between the
outer and inner regions.

As mentioned earlier, there is a
reasonable guess for the dual state: this would be the pure
state obtained by performing the Euclidean path integral
over the Riemann surface $\Sigma$ that is the conformal boundary of the
Euclidean 3-manifold that we glue to the Lorentzian spacetime  at $t=0$.
However, this appears at odds with the presence of a bulk
horizon. It would be interesting to clarify this and also
check the identification of the state
by computing in the CFT
the expectation value of the stress energy tensor in this state
and see if the results agree with our bulk computation.

One of the main reasons the black hole entropy has been so puzzling
is that classically black holes appear to be unique (they have
``no-hair'') so their phase space is zero dimensional.
In a typical quantum system the correspondence principle
relates the quantum states to the classical phase space
and the entropy of the system to the volume of phase space
in Planck units. Thus since the phase space for black holes
appears to be zero dimensional, they should not carry any entropy.
As was discussed earlier, however, the outer region of the wormholes
is isometric to the BTZ black hole. Thus one can view the
`wormhole' spacetimes with a single outer region
as `BTZ hair', where the `hair' is essentially
the non-trivial topology hidden behind the horizon.
It is thus natural to ask whether this classical phase space
can account for the entropy of the BTZ black hole
upon quantization\footnote{This \label{ftn}
question has been independently pursued by Alex Maloney \cite{Maloney}.}.
In other words, these spacetimes would then
be the semi-classical approximation of the underlying
black hole microstates.
This is similar in spirit to the fuzzball proposal
(whose relation to holography was discussed
extensively in the review \cite{Skenderis:2008qn})
although here the geometries counted contain horizons and singularities.

Let us outline how one would do such a computation.
We have seen that these spacetimes are uniquely specified by a Riemann surface
with one boundary and the mass of the BTZ black hole
is determined by one of the moduli of the Riemann surface.
Thus the classical phase space is the moduli space of
Riemann surfaces of arbitrary genus with a single fixed
modulus, corresponding to the length of the horizon (in other words
the BTZ mass parameter), which is the only parameter accessible to an
observer outside of the horizon. More precisely, if one uses
the Fenchel-Nielsen coordinates on the Teichm\"uller
space (described in detail in appendix \ref{app:FN})
the restriction to a fixed BTZ mass amounts to
considering a codimension one hypersurface in  Teichm\"uller
space. This hypersurface is invariant under the mapping class group
and therefore directly descends to the moduli space.
The complete phase space is then
the union of these hypersurfaces for different
genera. Classically,
the volume of this phase space is infinite and one should
proceed by geometric quantization.
One can readily compute the symplectic form
on the covariant space following
\cite{Crnkovic:1986ex,Crnkovic:1986be,Lee:1990nz}
and proceed to quantize. It would be interesting to carry out
this computation. The explicit form of the metric derived
in the appendix should facilitate this.

\section*{Acknowledgments}
We would like to thank Alex Maloney,
Jan Smit and Erik Verlinde for discussions. KS acknowledges support
from NWO via a VICI grant.

\appendix

\section{Coordinate systems}
\label{app:coordinatesystems}

The description of the wormholes in section \ref{sec:lorentzian} as a
quotient $\wAdst/\hat \Gamma$ is precise but rather abstract. This
appendix presents a metric description of the wormholes, building on \cite{mythesis}. More details are presented in \cite{myphdthesis}. Concretely, this
description consists of covering the spacetime with a set of charts
for which the coordinates have natural ranges. We then show that on
each of these charts we can put
an explicit metric, which features several
natural parameters that describe the local geometry (similar to the
mass $M$ for a BTZ metric). We will show that one may arrive at a
complete description of the spacetime by combining the parameters from
all the charts plus specifying some combinatorial data, which can be
combined in a single labelled fatgraph. An example of such a fatgraph
is given in figure \ref{fig:graph}, which completely describes a
spacetime with the topology sketched in figure \ref{fig:sketch}. The
particular parameters that will appear in the metric are very similar
to Fenchel-Nielsen coordinates on Teichm\"uller space, so we begin
with a review of these coordinates.

\subsection{Fenchel-Nielsen coordinates} \label{app:FN}
In this section we review the definition of the Fenchel-Nielsen coordinates on the Teichm\"uller space of Riemann surfaces of genus $g$ with $m > 0$ circular boundaries (and no punctures). As we discussed in the main text, all such Riemann surfaces are quotients of the upper half plane, from which they all inherit a canonical metric of constant negative curvature.

\begin{figure}
\centering
\psfrag{l1}{$l_1$}
\psfrag{l2}{$l_2$}
\psfrag{l3}{$l_3$}
\psfrag{l4 t4}{$l_4\, \, t_4$}
\psfrag{l5 t5}{$l_5 \, \,t_5$}
\psfrag{l6 t6}{$l_6 \, \, t_6$}
\psfrag{l7}{$l_7$}
\psfrag{l8}{$l_8$}
\psfrag{t1}{$t_1$}
\psfrag{t2}{$t_2$}
\psfrag{t3}{$t_3$}
\psfrag{t4}{$t_4$}
\includegraphics[width=10cm]{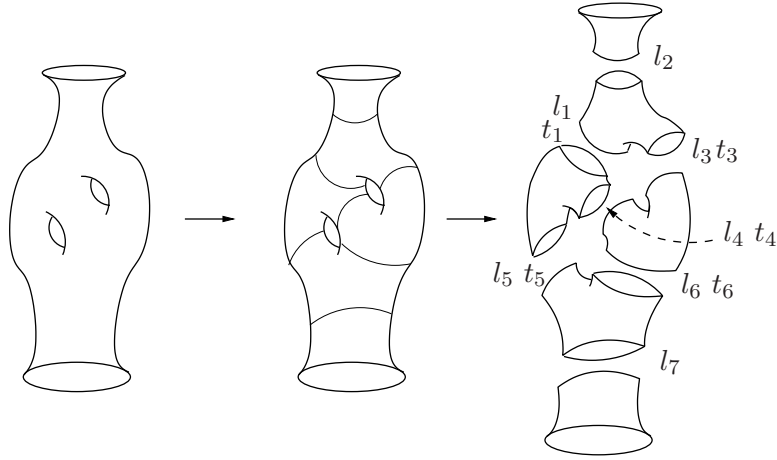}
\caption{\label{fig:fenchelnielsen}Defining Fenchel-Nielsen coordinates on a Riemann surface. We cut the Riemann surface into pairs of pants along simple closed geodesics and assign lengths $l_i$ to all the edges of every pants plus a twisting parameter $t_j$ for every gluing involving two pairs of pants.}
\end{figure}

It can be shown that in this metric there is precisely one smooth periodic geodesic corresponding to every nontrivial primitive loop on the surface. After a little counting one finds that one can pick a maximum of $3g - 3 + 2 m$ of such periodic geodesics that do not intersect each other, see figure \ref{fig:fenchelnielsen} for an example. We then \emph{cut} the Riemann surface along these geodesics, \ie we remove these geodesics from the surface. This leaves us with $2g - 2 + m$ disconnected so-called `pairs of pants', that is Riemann surfaces of genus $0$ with three circular boundary components, as well as $m$ annuli. The annuli correspond to the regions on the Riemann surface between a periodic geodesic that is retractable into a boundary component and the boundary component itself.

The Fenchel-Nielsen coordinates are now based on the idea that we can reconstruct the complete Riemann surface from this collection of pairs of pants and annuli, provided we also specify how to glue these `building blocks' together. Therefore, we can define coordinates on the Teichm\"uller space of Riemann surfaces of the given type by specifying enough data to first of all construct the pairs of pants and annuli that make up the original surface, plus some rules on how to glue them together.

Let us begin with a description of the individual pairs of pants and annuli. Using some simple hyperbolic geometry, see for example \cite{imayoshitaniguchi}, one finds that the pairs of pants are completely described by only three real moduli which one may take to be the strictly positive lengths of the periodic geodesics along which we made the cuts. A similar statement is true for the annuli: these are completely specified by giving the length of the periodic geodesic as well. Since we cut along $3g -3 + 2m$ periodic geodesics, we find that we can reconstruct the individual pairs of pants and annuli by the specification of precisely $3g -3 + 2m$ strictly positive lengths.

\begin{figure}
\centering
\psfrag{p}{$p$}
\psfrag{pp}{$p'$}
\psfrag{q}{$q$}
\psfrag{t}{$t$}
\includegraphics[width=11cm]{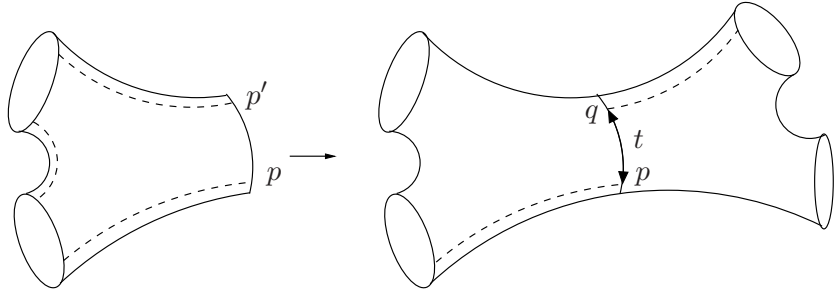}
\caption{\label{fig:twist}The twist parameter $t$ is defined by the angle between two points $p$ and $q$ that lie at the intersection of the dashed geodesics with a boundary circle.}
\end{figure}

Next, we have to specify the way in which the various components are glued together. More specifically, we have to specify the angle that the various components have to be twisted with before we perform the gluing. Notice that these angles are actually only relevant when we glue two pairs of pants together, since twisting an annulus is an isometry. The angles are defined as follows, see figure \ref{fig:twist}. On every pair of pants we may define three distinguished geodesics, namely the shortest non-intersecting geodesics that run from one boundary circle to another. A given boundary circle of the pants intersects with two of these geodesics, say at the points $p$ and $p'$. (Figure \ref{fig:twist} is drawn slightly distorted since these points actually lie diametrically opposite of each other. This follows from a reflection isometry of the pair of pants whose fixed points are precisely the three geodesics we just defined.) Following the same reasoning on the other pair of pants we find two more points, say $q$ and $q'$, on this boundary circle. The twist parameter describing the gluing is now precisely the angle between, say $p$ and $q$, on the boundary circle.\footnote{A shift of $2\pi$ in the angles corresponds to an element of the mapping class group and therefore to two different points in Teichm\"uller space. Strictly speaking, therefore, these angles take values in $\Real$ in order to properly parametrize the Teichm\"uller space. We will be rather loose in this distinction.}

Since we cut along $3g - 3 + 2m$ geodesics, we have as many gluings to perform. For precisely $m$ of these we glue annuli to pairs of pants, which leaves us with $3g - 3 + m$ gluings between pairs of pants for which we need to specify an angle. Adding these to the $3g - 3 + 2m$ lengths precisely gives the required number of $6g - 6 + 3m$ parameters. Indeed, it can be shown that these lengths and angles provide good coordinates that cover the Teichm\"uller space of Riemann surfaces of the given type, which is therefore isomorphic to $(\Real^+)^{3g -3+ 2m} \times \Real^{3g - 3 + m}$. This is then the Fenchel-Nielsen description of the Teichm\"uller space.

\subsection{Construction of the charts}
\begin{figure}
\centering
\psfrag{f}{$\phi$}
\psfrag{r}{$r$}
\psfrag{a}{(a)}
\psfrag{b}{(b)}
\psfrag{c}{(c)}
\includegraphics[width=11cm]{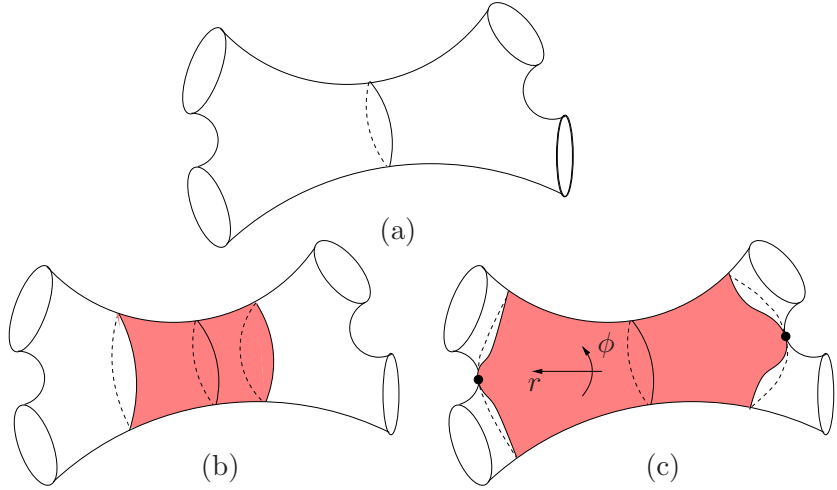}
\caption{\label{fig:charts}(a) Charts are defined around a closed periodic geodesic on the Riemann surface. (b) We begin by thickening this geodesic to obtain a cylinder. (c) We extend the cylinder as far as possible, until the bounding circles just touch, in this case on the black dots. We define coordinates $(r,\phi)$ as indicated, as well as a third time coordinate which is not shown.}
\end{figure}

The procedure to obtain our charts is sketched in figure \ref{fig:charts} and is described in words as follows. We first restrict ourselves to the $U = 0$ Riemann surface $S = H/\Gamma$.  Just as in the Fenchel-Nielsen description of the surface, we begin by picking a maximal set of $3g - 3 + 2m$ primitive periodic geodesics. We now consider one geodesic and `thicken' it, \ie we define a small cylindrical neighborhood around the geodesic. When we try to extend this `collar' further, eventually we might wrap another cycle and the cylinder will then start to overlap with itself. We then stop the thickening when the boundary circles just touch each other, as indicated in figure \ref{fig:charts}. In the cases where the periodic geodesic we consider is retractable into a boundary component we extend the thickening on that end all the way to this boundary. Except for the BTZ black hole, the other end of the cylinder is then never extendable to another boundary component and pinches as usual.

This procedure results in two types of cylindrical domains: those where both boundary circles are pinched on $S$, which we call `inner domains', and those where precisely one end extends to a boundary component, which we call `outer domains'. An inner domain covers part of two pairs of pants, whereas an outer domain covers an annulus and part of a pair of pants.

Notice that the inner and outer domains we define here are not precisely the inner and outer regions we defined in the main text. Namely, the inner and outer regions in the main text were separated by the horizons, whereas the outer domains we define here do extend beyond the horizons. The inner domains that we define here never cross the horizons and therefore lie entirely in what we called the inner region in the main text.

Consider now a single pair of pants. It intersects with precisely three (inner or outer) domains, namely those that are defined around each of its boundary circles. Of course, the domains overlap with each other on the pants but more importantly it can be shown that the \emph{entire} pair of pants is covered by these three domains. (This follows from direct computation using hyperbolic geometry, see \cite{myphdthesis} for details.) Since the domains also cover the annuli completely, it follows that the entire surface at $U=0$ is covered by these domains. Below, we will use these domains as the $U=0$ slice of analogously defined three-dimensional coordinate patches, which taken together cover the entire spacetime. We will then find a suitable coordinate system on these patches to complete our description of the wormholes.

\begin{figure}
\centering
\psfrag{r}{$r$}
\psfrag{f}{$\phi$}
\psfrag{la}{$l_\alpha$}
\psfrag{lb}{$l_\beta$}
\psfrag{glb}{$\gamma(l_\beta)$}
\includegraphics[width=9cm]{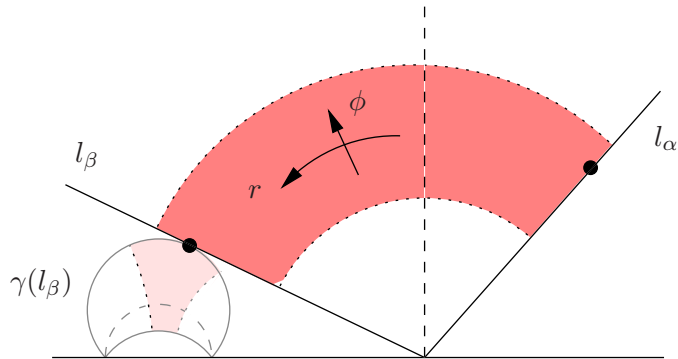}
\caption{\label{fig:chartinh}A lift to the upper half plane of the cylindrical region. The dotted lines bounding the darker region should be identified under the map $w \sim \lambda w$. Under $\gamma_\beta$, the region maps to the smaller, lighter shaded region on the left. The image $\gamma(l_\beta)$ of $l_\beta$ just touches $l_\beta$ at the indicated point. A similar thing happens on the right for $l_\alpha$, but we have not sketched the second image.}
\end{figure}

To define more precisely the inner and outer domains let us lift them to the universal cover $H$ of $S$, where we will use a complex coordinate $z$. Consider one of the periodic geodesics around which we defined a chart. We assume that on $H$ the homotopy class of the periodic geodesic is generated by the identification
\be
\label{eq:simplemap}
\gamma: z \mapsto \lambda z\,,
\ee
which can always be realized using the conjugation freedom of $\Gamma$. If $z = x+ iy$, then the periodic geodesic lifts to the line $x  = 0$. The corresponding lift of the cylindrical neighborhood around it is a region $D$ given by
\be
D: - \beta y < x < \alpha y\,,
\ee
for some positive real $\alpha$ and $\beta$ (which are given in terms
of the Fenchel-Nielsen parameters that fix the geometry of the pairs
of pants, as it will become clear from the analysis below). For an inner domain
$\alpha$ and $\beta$ are finite whereas for an outer domain either
$\alpha$ or $\beta$ are equal to $+\infty$ and the domain extends all
the way to the boundary. A region $D$ with finite $\alpha$ and
$\beta$, \ie corresponding to an inner domain, is sketched in figure
\ref{fig:chartinh}. The lines $l_\alpha$ and $l_\beta$, given by
$\alpha y = x$ and $- \beta y = x$ respectively, determine the
bounding circles of the cylindrical neighborhood. We have deliberately
chosen the shape of these bounding circles such that they lift to
straight lines on $H$ which are called \emph{hypercycles}. (Recall
that geodesics on $H$ are either semicircles that are orthogonal to
the real axis or straight vertical lines; hypercycles, on the other
hand, are straight lines or circle segments that end on the real axis
but not at a right angle. Examples are $l_\alpha$, $l_\beta$ and
$\gamma(l_\beta)$ in figure \ref{fig:chartinh}.) As we mentioned
above, the cylindrical neighborhood is `maximally extended' in the
sense that its bounding circles on $S$ touch themselves somewhere on
$S$. Correspondingly, there must exist $\gamma_\alpha, \gamma_\beta
\in \Gamma$ that map $l_\alpha,l_\beta$ to circle segments that just
touch $l_\alpha,l_\beta$ on $H$. We have sketched this in figure
\ref{fig:chartinh}.

The cylinder can now be extended to a region on the full three-dimensional wormhole geometry. We first extend the action of the isometry \eqref{eq:simplemap} to the Poincar\'e patch:
\be
\label{eq:hatgamma}
\hat \gamma: (t,x,y) \sim \lambda (t,x,y)\,,
\ee
and then extend the domain $D$ to an invariant domain $\hat D$ in the full three-dimensional geometry. For inner domains it is defined as
\be
\label{eq:hatD}
\hat D:  - \beta \sqrt{y^2 - t^2} < x < \alpha \sqrt{y^2 - t^2} \,,
\ee
with $y^2 - t^2 > 0$. For outer domains either $\alpha$ or $\beta$ are equal to $+\infty$ and correspondingly
there is no restriction on the sign of $y^2 - t^2$ when $x > 0$ or $x < 0$, respectively. On that end the outer domain extends all the way to the conformal boundary of the spacetime.

We note that the region with $-t^2 + x^2 + y^2 \leq 0$ has to be excluded because it lies within the future and past lightcone of the origin, which is a fixed point of the isometry \eqref{eq:hatgamma}. One may check that \eqref{eq:hatgamma} indeed leads to closed timelike or lightlike curves in this region. There are other excluded regions that are bounded by lightcones with their vertex at the point at infinity but these are precisely the regions in \Adst that are not covered by the Poincar\'e coordinate system anyway.

Let us now sketch a proof for the covering of the entire spacetime by these domains. First of all, notice that the future and past Cauchy development of the $t=0$ slice (which we will call $C$) is covered by the part of the Poincar\'e coordinate system with $y^2 - t^2 > 0$. Then from \eqref{eq:hatD} we see that the inner charts all lie within this domain. The domain $C$ can be foliated with slices of constant $U = t/y$ on which the quotient group $\hat \G$ acts just as on the initial $U=0$ surface. The covering of $C$ then follows straightforwardly from the fact that the $U=0$ surface is covered.

However, the wormholes 
are not globally hyperbolic and a part of the wormhole spacetime near the conformal boundary lies outside of $C$. To find the shape of this part of the spacetime we notice the following. Near the conformal boundary the spacetime has the form of an annulus times a time coordinate and when we move inward this annulus pinches just as in figure \ref{fig:charts}c. It follows from \eqref{eq:hatD} that
the pinching occurs either at $x=- \b \sqrt{y^2 - t^2} $ or at $x=\a \sqrt{y^2 - t^2} $ 
for some finite $\a, \b$. Either way this `pinching surface' must lift to a region with $y > |t|$  and therefore always lies entirely within $C$. It follows that the parts of the spacetime outside of $C$ must have the shape of an annulus times time. It is then easy to verify that these regions of the spacetime outside of $C$ can be described in Poincar\'e coordinates by starting with the region where $y < |t|$ and $x > 0$, excluding the lightcones where $-t^2 + x^2 + y^2 \leq 0$ and taking the quotient of the remainder with respect to the cyclic group generated by \eqref{eq:hatgamma}. Indeed, the domains so obtained are bounded by the lightlike surfaces $y = |t|$ that bound $C$, extend all the way to the conformal boundary $y=0$ and the action of the cyclic covering group guarantees that the quotient has the form of an annulus times time. These regions are by construction also completely covered by an outer domain and therefore indeed the entire spacetime is covered.

One may also explicitly verify that the coordinate systems on the inner and outer domains as given in \eqref{eq:newinnercoords} and \eqref{eq:newoutercoords} below are everywhere well-defined on these domains.

\subsubsection{Coordinate systems on inner domains}
We may now define new coordinates on the three-dimensional domains $\hat D$. For inner domains we define a coordinate system $(\tau,r,\phi)$ via:
\begin{align}
\label{eq:newinnercoords}
\tanh(\tau) &= \frac{t}{y}\,, &
\mx r + \nx &= \frac{x}{\sqrt{y^2 - t^2}}\,, &
e^{2 \sqrt M \phi} &= - t^2 + x^2 + y^2\,,
\end{align}
with coefficients
\be \label{eq:in_par}
\begin{split}
e^{2 \pi \sqrt M} = \lambda \,, \qquad \qquad \mx + \nx = \alpha\,, \qquad \qquad \mx - \nx = \beta\,.
\end{split}
\ee
From \eqref{eq:hatgamma} and \eqref{eq:hatD} we find the coordinate ranges:
\be
\label{eq:coordrangesinner}
\tau \in \Real\,,\qquad \qquad \phi \sim \phi + 2\pi\,,\qquad \qquad  r \in [-1,1]\,,
\ee
and the metric takes the form:
\be
ds^2 = \frac{1}{\cosh^2(t)}\Big( -dt^2 + \frac{\mx^2 dr^2}{(\mx r + \nx)^2 + 1} + M (1 + (\mx r + \nx)^2) d \phi^2 \Big)\,.
\ee
This metric already features several parameters $M,\mx,\nx$ which inform us about the geometry at least in this local patch. We can however introduce one more parameter which is related to the Fenchel-Nielsen twist described above.

\begin{figure}
\centering
\psfrag{f=0}{$\phi = 0$}
\psfrag{f=pi}{$\phi = \pi$}
\includegraphics[width=6cm]{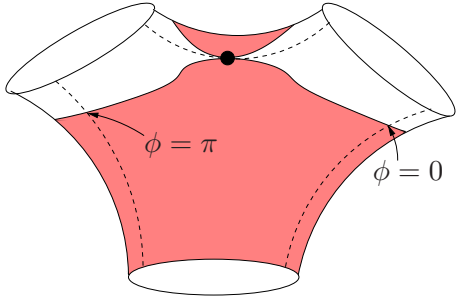}
\caption{\label{fig:phasephi}The phase in $\phi$ can be fixed by letting $\phi = 0$ correspond to the intersection of the boundary of the chart with the unique shortest length geodesic (dashed line) between the two boundary circles of the corresponding pair of pants.}
\end{figure}

To find this parameter, let us begin by considering one edge of a particular chart, say at $r=+1$. As indicated in figure \ref{fig:phasephi}, such an end lies at a pair of pants that is used in the Fenchel-Nielsen description of the surface. We described before that there are three shortest geodesics on this pair of pants that run between the three boundary components, see figure \ref{fig:twist}. As we sketched in figure \ref{fig:phasephi}, two of these geodesics intersect the boundary circle of the chart. We can now shift $\phi$ such that one of these intersection points corresponds to $\phi = 0$ and from the aforementioned reflection symmetry it follows that the other one automatically lies at $\phi = \pi$. (It can also be shown that the third of these geodesics precisely touches the boundary of the charts at the pinching point which is indicated by the black dot in figure \ref{fig:phasephi}.)

After having implemented this shift at the side $r = +1$ we find that the corresponding points at the side $r = -1$, which lie on another pair of pants, generally lie at a value $\phi = \phi_0$ and $\phi = \phi_0 + \pi$, all modulo $2 \pi$. In fact, the angle $\phi_0$ is precisely the Fenchel-Nielsen twist coordinate (denoted $t$ above) that is associated to the gluing. We can make this twist explicit in the metric by introducing a new `twisted' coordinate $\psi$ given by:
\be
\label{eq:newinnerphi}
\exp(\sqrt M \psi  - k) = \exp(\sqrt M \phi) f(\mx r + \nx, \chi)
\ee
with
\be
f(\rho,\chi) = \frac{\rho \sin(\chi) + \sqrt{\rho^2 + \cos^2(\chi)}}{\sqrt{\rho^2 + 1}}\,.
\ee
This coordinate transformation features two new parameters $k$ and $\chi$. If they are chosen such that
\be
e^{-k} = f(\mx + \nx,\chi) = e^{\sqrt M \phi_0} f(- \mx + \nx,\chi)\,,
\ee
then the aforementioned distinguished points are given by $\psi = 0$ and $\psi = \pi$ on both sides. The coordinate range of $\psi$ is the same as $\phi$, so $\psi \sim \psi + 2 \pi$. The parameter $\chi$ now shows up explicitly in the metric, which takes the form:
\be
\begin{split}
ds^2 &= \frac{1}{\cosh^2(t)}\Big( -dt^2 + \frac{\mx^2 dr^2}{(\mx r + \nx)^2 + \cos^2(\chi)} + M \big(1 + (\mx r + \nx)^2\big) d \psi^2 \\ &\qquad \qquad - \frac{2 \mx \sqrt M \sin(\chi)}{\sqrt{(\mx r + \nx)^2 + \cos^2 (\chi)}} \, d\psi d r \Big)\,.
\end{split}
\ee
This is the final metric on the inner chart. The four different parameters $M,\mx,\nx,\chi$ appearing in it inform us about some `local' geometrical aspects of the spacetime. Namely, the periodic geodesic around which we defined the chart lies at the point $r = - \nx/\mx$ and has length $2\pi \sqrt M$. The angle $\chi$ reflects the twisting of the pairs of pants with respect to each other and the parameters $\mx$ and $\nx$ are related to the shapes of these pairs of pants: for example, the distance between the periodic geodesic and the pinched hypercycle at $r = 1$ is
\[
| \ln\Big( \frac{\mx + \nx + \sqrt{(\mx+\nx)^2 + 1}}{\nx + \sqrt{\nx^2 + 1}}\Big) |\,,
\]
and the distance to the hypercycle at $r = -1$ has the same form with the replacement $\mx \to -\mx$.

\subsubsection{Coordinate systems on outer domains}
For the outer domains we may always conjugate $\Gamma$ such that $\a = \infty$ and $\b$ is finite. We can then use a $(\tilde \tau, \r,\varphi)$ coordinate system defined as
\begin{align}
\label{eq:newoutercoords}
\tanh(\sqrt M \tilde \tau) &= \frac{t}{\sqrt{y^2 + x^2}}\,, &
\r &= \sqrt M \, \frac{x}{y}\,, &
e^{2 \sqrt M (\varphi - h) } &= - t^2 + x^2 + y^2\,,
\end{align}
where, as in (\ref{eq:in_par}),
\be
e^{2 \pi \sqrt M} = \lambda\,,
\ee
which is again related to the length of the periodic geodesic. The parameter $h$ shifts the coordinate $\varphi$ such that the aforementioned special points on the bounding circle lie again at $\varphi = 0$ and $\varphi = \pi$. We will not need the explicit value of $h$ below. The bounding circle itself is given by
\be
\frac{\cosh(\sqrt M \tilde \tau)\r}{\sqrt{\r^2 + M}} = - \frac{\b^2}{1 + \b^2}\,,
\ee
and the coordinate ranges are given by
\be
\label{eq:coordrangesouter}
\tilde \tau \in \Real\,,\qquad \qquad \varphi \sim \varphi + 2\pi\,,\qquad \qquad \frac{\cosh(\sqrt M \tilde \tau)\r}{\sqrt{\r^2 + M}} > - \frac{\b^2}{1 + \b^2}\,.
\ee
The radial boundary of the spacetime lies at $\r \to \infty$. The metric takes the form:
\be
\label{eq:outermetric}
ds^2 = \frac{\rho^2 + M}{\cosh^2 (\sqrt M \tilde \tau)}(-d\tilde \tau^2 + d\varphi^2) + \frac{d\rho^2}{\rho^2 + M}\,.
\ee
Notice that these coordinate systems extend beyond the future and past horizons, which lie at the surfaces $x = |t|$ or
\be
\r = \sqrt M |\sinh(\sqrt M \tilde \tau)|\,.
\ee
The metric in the region outside of these horizons (which we called the outer region in the main text) can be put back in BTZ form \eqref{eq:btzmetric} by the coordinate transformation:
\be
r^2 = \frac{\r^2 + M}{\cosh^2(\sqrt M \tilde \tau)}\,, \qquad \tanh(\sqrt M t) = \sqrt{1 + M/\r^2} \tanh(\sqrt M \tilde \tau)\,,\qquad \phi = \varphi.
\ee
Notice that the parameter $M$ in \eqref{eq:outermetric} agrees with the BTZ mass $M$.

\subsection{Parameters} \label{sec:parameters}

The above charts can be combined to cover the wormhole spacetime
completely. More specifically, for a wormhole of genus $g$ and with
$m$ boundaries, we can cover the entire spacetime with $3g -3 + m$
inner charts plus $m$ outer charts. For every inner chart
we have four parameters,
$M,\mx,\nx,\chi$, and for every outer chart we have a single parameter
$M$. As we showed above, the angles $\chi$ and the parameters
$M$ are directly related to the Fenchel-Nielsen twists and length
parameters associated to the periodic geodesics and should therefore
completely determine the surface. The remaining $\mx$ and $\nx$
parameters are therefore expressable in terms of those.

The precise relation takes the following form. Consider a pair of pants in the surface. In our description of the surface it is covered by three (inner or outer) charts, in fact it is already completely covered by only half of each of these three charts. Suppose now that chart number 3 is an inner chart (with parameters $\mx_3,\nx_3,M_3,\chi_3$) and that it is the half with $r > 0$ that lies on the pair of pants under consideration. Denote the $M$ parameters in the other two charts as $M_i$ with $i \in \{1,2\}$. One then finds the relation:
\be \label{eq:m+n}
\mx_3 + \nx_3 = \frac{\sqrt{C_1^2 + C_2^2 + 2 C_1 C_2 C_3}}{\sinh(\pi \sqrt M_3)}\,,
\ee
with $C_i = \cosh(\pi \sqrt M_i)$. This relation follows from a straightforward computation in the upper half plane using hyperbolic geometry. A similar relation can be found at the other side of chart number 3, which has $r < 0$ and lies on another pair of pants. Namely, using the parameters $M'_1$ and $M'_2$ of the two other charts on that pair of pants we find:
\be \label{eq:n-m}
- \mx_3 + \nx_3 = \frac{\sqrt{C'{}_1^2 + C'{}_2^2 + 2 C'_1 C'_2 C_3}}{\sinh(\pi \sqrt M_3)}\,,
\ee
with $C'_i = \cosh(\pi \sqrt{M'_i})$. Using these formulae, we can determine all the $\mx,\nx$ parameters in the inner charts if we are only given the $M$ parameters in every chart. This reduces the number of independent parameters to two per inner chart and still one per outer chart, just as for the Fenchel-Nielsen description of the surface.

\subsection{Fatgraph description}
To completely specify the spacetime we need to specify both the parameters and the way the charts are glued together. This combinatorial data can be nicely summarized in an oriented trivalent fatgraph as shown in figure \ref{fig:graph}. (In the usual Fenchel-Nielsen description of the surface this combinatorial data is implicitly specified, for example by using a reference surface. The description given below, on the other hand, explicitly fixes the required combinatorial data and it is then no longer necessary to use a reference surface.)

The data in the fatgraph is translated to the coordinate systems as follows. Every edge represents a periodic geodesic and therefore a chart. Every vertex represents a pair of pants. The orientation of the edges indicates the direction of increasing $r$ (and by convention always points outward for outer charts), and the `fattening' is necessary to indicate how three charts come together on a pair of pants. If we add to this fatgraph two parameters $M,\chi$ for every interior edge of the graph and a single parameter $M$ for every outer edge, then the wormhole spacetime is completely specified.

At this point we should note that there are two discrete ambiguities in the above definitions of the coordinates $\psi$ and $\varphi$ on the inner and outer charts that we have not yet dealt with. Although these ambiguities do not affect the metric or the coordinate ranges given above, they will affect the transition functions below and therefore they should be fixed.

The first ambiguity involves the direction of increasing $\psi$ and $\varphi$. With the fatgraph description this can be easily fixed by fixing the handedness of the $(r,\psi)$ or $(r,\varphi)$ coordinate system to be the same in every chart.

\begin{wrapfigure}{L}{6cm}
\centering
\psfrag{0}{$0$}
\psfrag{pi}{$\pi$}
\psfrag{phi}{$\psi$}
\psfrag{r}{$r$}
\psfrag{r=+1}{$r=+1$}
\psfrag{r=-1}{$r=-1$}
\includegraphics[width=5cm]{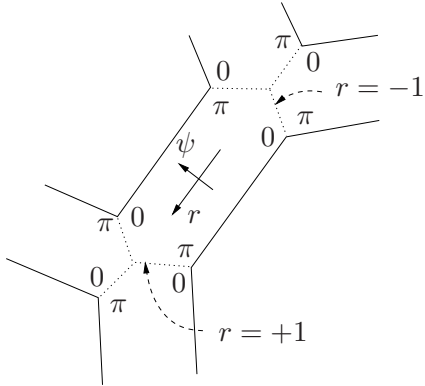}
\caption{\label{fig:graph_3} Fixing the ambiguities in the definition of $\psi$ and $\varphi$.}
\end{wrapfigure}

The second ambiguity is the fact that we have only $\psi$ or $\varphi$
up to an overall shift by $\pi$. To see this, recall that we decided
that the point $\psi = 0$ or $\varphi = 0$ would correspond to one of
the distinguished points on the boundary circle (sketched in figure
\ref{fig:phasephi}) and by the reflection isometry the other point
would then be at $\psi = \pi$ or $\varphi = \pi$. We however did not
yet specify \emph{which} point we chose to be at $0$ and which one at
$\pi$. This ambiguity can be fixed from the fatgraph. We first demand
that at an overlap between two charts the point where $\psi = 0$ on
one chart corresponds to $\psi = \pi$ on the other chart (and
similarly for $\varphi$), as indicated in figure
\ref{fig:graph_3}. Furthermore, for an inner chart we should
alternately associate $\psi =0$ and $\psi = \pi$ to the four corners
of the corresponding edge in the fatgraph, which is indicated in
figure \ref{fig:graph_3} as well. This fixes the ambiguity up to an
overall shift of $\psi$ or $\varphi$ with $\pi$ in all charts at the
same time, which is however irrelevant for the description of the
manifold.

\subsection{Transition functions}
With all the ambiguities fixed, we may proceed to define transition functions on the overlap between two different charts. These follow from the coordinate transformations \eqref{eq:newinnercoords} and \eqref{eq:newoutercoords} plus the explicit form of the elements of $\hat \Gamma$ in Poincar\'e coordinates (which can be deduced from \eqref{eq:adstaction} and \eqref{eq:poincarecoords}).

An important subtlety is that we find different transition functions depending on the gluings and the orientations of the charts. For example, if we consider the vertex in figure \ref{fig:graph} where we may go from chart 2 to chart 3 or chart 4, we find different transition functions because we turn `right' at the vertex if we go to chart 3, whereas we turn `left' if we go to chart 4. As another example, the transition functions between chart 2 and chart 3 (on both vertices) are different from those between chart 5 and chart 6 because (again on both vertices) the orientation of chart 3 and chart 6 are not the same. When we define the transition functions below we will have to take into account these different possibilities.

In the transition functions we will not use the `twisted' coordinate
$\psi$ defined in \eqref{eq:newinnerphi}. Instead, we will use the
coordinate $\phi$ which agrees with $\psi$ at the bounding
circle of the chart where we define the transition function. Of
course, it is not hard to compose the transition functions with
\eqref{eq:newinnerphi} and its inverse, or a similar function when the
transition takes place at $r = -1$.

\subsubsection{Transitions between two inner charts}
\begin{figure}
\centering
\psfrag{1}{$1$}
\psfrag{2}{$2$}
\psfrag{3}{$3$}
\psfrag{4}{$4$}
\psfrag{5}{$5$}
\begin{tabular}{cc}
\includegraphics[width=6cm]{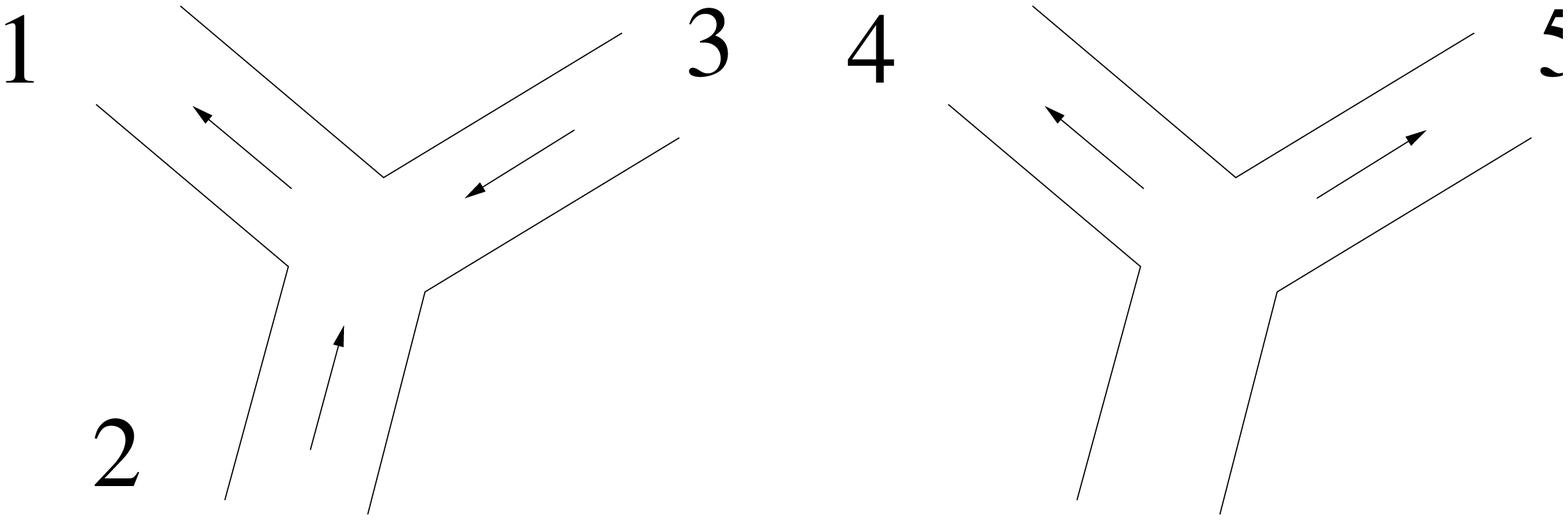}
&
\begin{tabular}{ccccc}
& $\epsilon$ & $\epsilon'$ & $d$ & $d'$\\ \hline
$1 \to 2$ & $-1$ & $+1$ & 0     & $\pi$ \\
$1 \to 3$ & $-1$ & $+1$ & $\pi$ & 0 \\
$2 \to 1$ & $+1$ & $-1$ & $\pi$ & 0 \\
$2 \to 3$ & $+1$ & $+1$ & 0     & $\pi$  \\
$3 \to 1$ & $+1$ & $-1$ & 0     & $\pi$  \\
$3 \to 2$ & $+1$ & $+1$ & $\pi$ & 0 \\
$4 \to 5$ & $-1$ & $-1$ & $\pi$ & 0 \\
$5 \to 4$ & $-1$ & $-1$ & 0     & $\pi$  \\
\end{tabular}
\end{tabular}
\caption{\label{fig:transitions}Possible transitions between inner charts. The transition functions are by definition always taken from unprimed to primed coordinate systems: for example, in the first line the unprimed coordinates in \eqref{eq:transition1to1} are the coordinates in chart 1 and the primed coordinates are those of chart 2.}
\end{figure}

The complete set of possibilities for the transitions between two inner charts is depicted in figure \ref{fig:transitions}. As one may expect, the transition functions are \emph{almost} the same for either one of these possibilities and it is convenient to give them in a general form with certain parameters $\epsilon$, $\epsilon'$, $d$ and $d'$ whose value depends on these possibilities and is given in the table in figure \ref{fig:transitions}. Using these parameters, one finds for the transition functions,
\begin{equation}
\label{eq:transition1to1}
\begin{split}
t' &= t \\
- \epsilon' (\mx' r' + \nx') &= \cosh(A)\epsilon (\mx r+\nx) - \sinh(A) \sqrt{(\mx r+\nx)^2 + 1} \cosh(\epsilon \sqrt M (\phi - d)) \\
e^{2 \epsilon' \sqrt {M'} (\phi' - d')} &= \frac{\epsilon(\mx r+\nx) - \sqrt{(\mx r+\nx)^2 + 1} \cosh(\epsilon\sqrt M (\phi - d) - g)}{\epsilon(\mx r+\nx) - \sqrt{(\mx r+\nx)^2 + 1}\cosh(\epsilon \sqrt M (\phi - d) + g)}
\end{split}
\end{equation}
with
\be
\cosh(A) = \frac{\cosh(\pi \sqrt{M})\cosh(\pi \sqrt{M'}) + \cosh(\pi \sqrt{M''})}{\sinh(\pi \sqrt{M})\sinh(\pi \sqrt{M'})}
\ee
and
\[
\sinh(A) \sinh(g) = 1.
\]
Here $M$ and $M'$ denote mass parameters in the metric on the unprimed and the primed chart between which we define the transition functions, and $M''$ denotes the mass parameter from the metric of the third chart that joins this vertex. We therefore have to inspect the metric of all three charts at the vertex in order to obtain the transition functions between only two of these charts.

Notice that the transition functions are not automatically periodic in $\phi$ or $\phi'$; they are in fact only valid for $\phi,\phi' \in [0,2\pi)$. Of course, this is by no means a restriction as this is sufficient to cover the entire chart. The other boundaries of the domain of validity of the transition functions are obtained from the coordinate ranges \eqref{eq:coordrangesinner}. For example, substituting $r' = 1$ in the second equation of \eqref{eq:transition1to1} one finds an equality involving $r$ and $\phi$ which defines the boundary of the domain of definition of the transition functions.

\subsubsection{Transitions involving outer charts}
\begin{figure}
\centering
\psfrag{1}{$1$}
\psfrag{2}{$2$}
\psfrag{3}{$3$}
\psfrag{4}{$4$}
\psfrag{5}{$5$}
\psfrag{6}{$6$}
\begin{tabular}{cc}
\includegraphics[width=6cm]{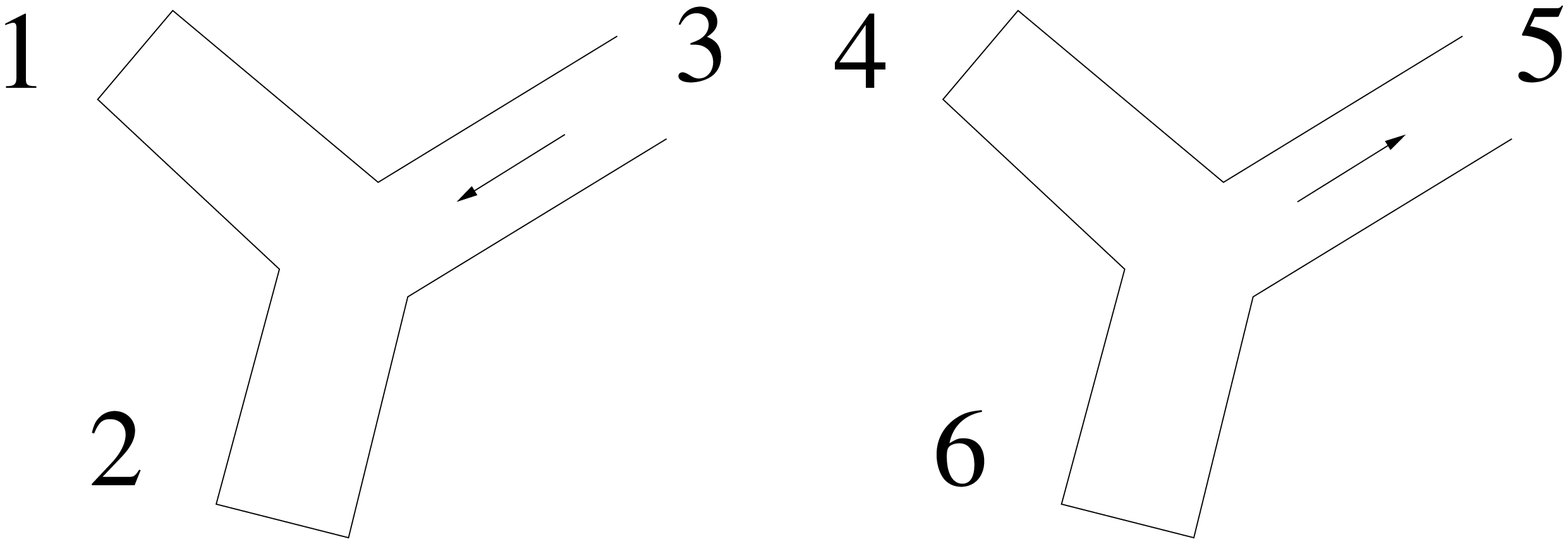} &
\begin{tabular}{ccccc}
&  &  & $f$ & $f'$\\ \hline
$1 \to 2$ & & & 0     & $\pi$ \\
$2 \to 1$ & & & $\pi$ & 0 \\
& & & & \\
& $\epsilon$ & & $d$ & $f'$\\ \hline
$3 \to 1$ & $+1$ & & 0     & $\pi$ \\
$3 \to 2$ & $+1$ & & $\pi$ & 0 \\
$5 \to 4$ & $-1$ & & 0     & $\pi$ \\
$5 \to 6$ & $-1$ & & $\pi$ & 0  \\
& & & & \\
& & $\epsilon'$ & $f$ & $d'$\\ \hline
$1 \to 3$ & & $+1$ & $\pi$ & 0  \\
$2 \to 3$ & & $+1$ & 0     & $\pi$ \\
$4 \to 5$ & & $-1$ & $\pi$ & 0 \\
$6 \to 5$ & & $-1$ & 0     & $\pi$  \\
\end{tabular}
\end{tabular}
\caption{\label{fig:transitions_2}Possible transitions involving outer charts. In this picture the charts 1,2,4 and 6 are outer charts and the charts 3 and 5 are inner charts. Conventions are as in figure \ref{fig:transitions}.}
\end{figure}

If the transitions involve outer charts we need the $(\tilde \tau, \r, \varphi)$ coordinate system. Since we always pick the $\r$ coordinate to increase towards the boundary there is no ambiguity on the orientation of this coordinate. We are however still left with the left/right ambiguity and correspondingly need a discrete parameter $f$ associated to every outer chart. For the transition functions between two outer charts we find,
\begin{gather}
\label{eq:transition2to2}
\begin{split}
&\rho' = - \sqrt{\frac{M'}{M}} \Big( \cosh(A) \rho + \sinh(A) \cosh(\sqrt M (\varphi -f)) \frac{\sqrt{\rho^2 + M}}{\cosh(\sqrt M \tilde \tau)} \Big) \\
&\sqrt{M} \tanh(\sqrt{M'}\tilde \tau') \sqrt{\rho'^2 + M'} = \sqrt{M'}\tanh(\sqrt M \tilde \tau) \sqrt{\rho^2 + M}  \\
&e^{2 \sqrt{M'} (\varphi' - f')} =
\frac{\rho \cosh(\sqrt M \tilde \tau) + \sqrt{\rho^2 + M} \cosh(\sqrt M (\varphi - f) - g)}{\rho \cosh(\sqrt M \tilde \tau) + \sqrt{\rho^2 + M} \cosh(\sqrt M (\varphi - f) + g)}
\end{split}
\end{gather}
with the possible values of $f$ and $f'$ given in figure \ref{fig:transitions_2} and the same values of $A$ and $g$ as before. The transition function on the second line is slightly implicit but it is straightforward to plug in the solution for $\r'$ of the first line and then solve for $\tilde \tau'$.

Similarly, between an inner and an outer chart we find,
\begin{gather}
\label{eq:transition1to2}
\begin{split}
&\rho' = \frac{\sqrt{M'}}{\cosh(t)} \Big( \cosh(A)\epsilon (\mx r+\nx) - \sinh(A) \sqrt{(\mx r+\nx)^2 + 1} \cosh(\epsilon \sqrt M (\phi - d)) \Big) \\
&\tanh(\sqrt{M'} \tilde \tau') \sqrt{\rho'^2 + M'} = \sqrt{M'} \tanh(t) \\
&e^{2 \sqrt{M'} (\varphi' - f')} =
\frac{\epsilon(\mx r+\nx) - \sqrt{(\mx r+\nx)^2 + 1} \cosh(\epsilon\sqrt M (\phi - d) + g)}{\epsilon(\mx r+\nx) - \sqrt{(\mx r+\nx)^2 + 1}\cosh(\epsilon \sqrt M (\phi - d) - g)}
\end{split}
\end{gather}
and conversely,
\begin{gather}
\label{eq:transition2to1}
\begin{split}
&\sqrt M \tanh(t') = \tanh(\sqrt M \tilde \tau) \sqrt{\rho^2 + M} \\
&\frac{\epsilon' (\mx'r'+\nx')}{\cosh(t')} = \sqrt{\frac{1}{M}} \Big( \cosh(A) \rho + \sinh(A) \cosh(\sqrt M (\varphi -f)) \frac{\sqrt{\rho^2 + M}}{\cosh(\sqrt M \tilde \tau)} \Big) \\
&e^{2 \epsilon' \sqrt{M'} (\phi' - d')} =
\frac{\rho \cosh(\sqrt M \tilde \tau) + \sqrt{\rho^2 + M} \cosh(\sqrt M (\varphi - f) + g)}{\rho \cosh(\sqrt M \tilde \tau) + \sqrt{\rho^2 + M} \cosh(\sqrt M (\varphi - f) - g)}
\end{split}
\end{gather}
Again, these transition functions are not obviously periodic in $\phi$ and $\varphi$ are are only valid in the interval $[0,2\pi)$ and the other boundaries are again found by inserting the coordinate ranges \eqref{eq:coordrangesinner} and \eqref{eq:coordrangesouter} in the transition functions. One may again compose the transition functions with \eqref{eq:newinnerphi} and its inverse to obtain the transition functions for the twisted coordinate $\psi$ on the inner charts.

\section{Eternal black holes and filled tori}
\label{app:fillings}

In this appendix we discuss the genus 1 handlebodies. We show that the
`$SL(2,\mathbb Z)$ family' of black holes cannot be used
in the real-time gauge/gravity prescription as the bulk filling of a
vertical segment of the QFT contour because the matching
conditions lead to a complex Lorentzian metric (and therefore $\vev{T_{ij}}$ does not satisfy the correct reality conditions, either).

Consider a Euclidean field theory on a torus with modular parameter
$\tau = \tau_1 + i \tau_2$. Without loss of generality we can pick
the circle given by $z \sim z + 1$ as the spatial circle along which
 we will cut open the Euclidean path integral and glue the Lorentzian
solutions. More
precisely, we will glue two Lorentzian cylinders to the lines $y = 0$
and $y = \tau_2 /2$, where $z = x+ iy$. As we discussed in \cite{us2},
$\tau_1$ is then $i$ times the angular momentum chemical potential,
but since  we are not
interested in rotating black holes here we will set $\tau_1$ to zero
throughout this appendix (it is straightforward
to generalize to $\tau_1 \neq 0$), so $\tau = i \tau_2$ is purely imaginary.

The torus so defined admits multiple bulk fillings, which are given by
the specification of a contractible cycle $z \sim z + a\tau + b$ with
$(a,b)$ two relatively prime integers. For
each of these fillings, one may obtain a complete Euclidean metric
which is locally $H^3$. After cutting the torus in half, we will
glue a Lorentzian bulk solution to the bulk hypersurface ending on the
lines $y=0$ and $y = \tau_2/2$. This hypersurface has the shape of an
annulus, except when $(a,b) = (0,1)$, when it consists of two
disks. In this case the matching Lorentzian solution is two segments
of thermal AdS. Notice also that for $(a,b) = (1,0)$ we obtain the
rotating BTZ black hole. To find the matching Lorentzian solutions in
the general case, we will first explicitly write down the Euclidean
bulk metric. We then investigate how the Lorentzian metric is
determined by the matching conditions.

\paragraph{Euclidean geometries\\}
Let us give a brief review of the possible fillings of the torus. We will again use the Poincar\'e coordinates $(\tau,x,y)$ defined in \eqref{eq:euclpoinc} on $H^3$, as well as the complex coordinate $w = x + i\tau$ on the boundary of $H^3$.  (Notice that the $\tau$ here is a coordinate and not the modular parameter of the torus. We henceforth exclusively use the coordinate $w$ so no confusion
should arise.) Any torus handlebody can be obtained as a quotient of $H^3$ by a cyclic group of identifications generated in Poincar\'e coordinates by:
\be
\label{eq:beta}
(w,y) \sim (e^{2 \pi i \beta}w ,|e^{2 \pi i \beta}|y)\,,
\ee
with $\beta = \beta_1 + i\beta_2$ a complex number.

Let us now compute the bulk metric when we use the complex boundary coordinate $z$ which has the natural periodicity $z \sim z + 1 \sim z + \tau$. We can do so using the map $J$ of section \ref{sec:holographic}. In this case, $J$ is a locally biholomorphic map from $\Complex$ rather than $H$, since the universal covering of the torus is $\Complex$ and not $H$. If the contractible cycle is given by $(a,b)$, the corresponding map $J : \Complex \to S^2$ is given by:
\be
J: z \mapsto w = e^{\alpha z}\,,
\ee
with $\alpha = 2\pi i (a \tau + b)^{-1}$. The identifications $z \sim z+1 \sim z+ \tau$ become
\be
\label{eq:identificationsw}
w \sim w e^{\alpha} \sim w e^{\alpha \tau}
\ee
which implies
\be
\label{eq:singleidentificationw}
w \sim e^{\alpha(c\tau + d)}w\,.
\ee
Now, since one trivially has that $w \sim e^{\alpha(a \tau + b)}w$, it
follows that the single identification \eqref{eq:singleidentificationw} is
equivalent to both identifications in \eqref{eq:identificationsw} provided
$ad - bc =1$.
Comparing \eqref{eq:singleidentificationw} with \eqref{eq:beta}, we then read off that
\be
\beta = \frac{c\tau + d}{a \tau + b}\,.
\ee
Following the same steps as in section \ref{sec:holographic}, we find that the bulk metric in the $z$ coordinate becomes
\be
ds^2 = \frac{d\rho^2}{\rho^2} + \frac{1}{\rho^2} | dz + \frac{\bar \alpha^2 \rho^2}{4} d \bar z|^2\,.
\ee
This metric is of the Fefferman-Graham form \eqref{eq:barefgmetric} and we can read off that the one-point function of the stress energy tensor is given by:
\be
\vev{T_{zz}} = \frac{\alpha^2}{2}\,,
\ee
which is again $-S[J]$, just as we found for the higher genus
handlebodies in section \ref{sec:holographic}.

\paragraph{Lorentzian geometry\\}

Let us now consider the continuation to the Lorentzian geometry. On
the boundary we cut open the Euclidean geometry along the circles
given by $y=0$ and $y = \tau_2/2$. In every case except thermal AdS
these circles are the boundary of a single annular region in the bulk
manifold.
Locally the
unique solution is simply given by analytic continuation. Using the
boundary lightcone coordinates $(u,v)$, we find the Lorentzian metric,
\be
ds^2 = \frac{d\rho^2}{\rho^2} + \frac{1}{\rho^2}
( du + \frac{\bar \alpha^2}{4} \rho^2 d v)
( dv + \frac{\alpha^2}{4} \rho^2du).
\ee
The periodicity for the boundary coordinates is $ (u,v) \sim (u + 1, v
+ 1)$, and $(u,v)$ are real whereas $\rho$ has the same range as
above. This metric is however \emph{complex} unless $\alpha^2$
is real, which only happens if either $a=0$ or $b=0$.
This is problematic both from the bulk and the holographic perspective.
In particular, the
expectation value of the dual stress energy tensor can be computed
using (\ref{eq:vevTgeneral}),
\be
\vev{T_{uu}} = \frac{1}{2} \a^2, \qquad \vev{T_{vv}} = \frac{1}{2} \bar{\a}^2,
\ee
and is complex, which cannot be the case for a hermitian operator.

\providecommand{\href}[2]{#2}\begingroup\raggedright\endgroup

\end{document}